\documentclass[prc,letterpaper,twocolumn,showpacs,showkeys,lengthcheck,tightenlines,
               nofootinbib,preprintnumbers,superscriptaddress]{revtex4-1} 

\pdfpageattr {/Group << /S /Transparency /I true /CS /DeviceRGB>>}

\usepackage{newtxtext,newtxmath}

\usepackage{titlesec}
\usepackage{dcolumn}
\usepackage{bm}
\usepackage{amsmath,amssymb,amsfonts}

\usepackage{graphicx}

\usepackage{color}
\usepackage{bbold}

\usepackage{slashed}
\usepackage[svgnames]{xcolor}
\usepackage[english]{babel}
\usepackage{blindtext}
\usepackage{microtype}
\usepackage{tikz}

\usepackage{ifpdf}
\usepackage[utf8]{inputenc}

\def\a{\alpha}
\def\b{\beta}

\def\g{\gamma}

\def\s{\sigma}

\def\hs{\hspace}

\def\no{\nonumber}

\def\lf{\left}
\def\rg{\right}

\newcommand{\vect}[1]{\boldsymbol{#1}}
\newcommand{\ph}[1]{\phantom{#1}}

\graphicspath{{.}{figures/}} 

\titlespacing{\section}{4pt}{10pt plus 4pt minus 2pt}{8pt plus 2pt minus 2pt}
\titlespacing{\subsection}{0pt}{12pt plus 4pt minus 2pt}{8pt plus 2pt minus 2pt}

\begin{document}

\title{The Sea of Quarks and Antiquarks in the Nucleon: a Review}

\author{D. F. Geesaman}
\affiliation{Argonne Associate of Global Empire, LLC, Argonne National Laboratory, Argonne, IL 60439}

\author{P. E. Reimer}
\affiliation{Physics Division, Argonne National Laboratory, Argonne, Illinois 60439, USA}

\begin{abstract}
  The quark and gluon structure of the proton has been under intense experimental and theoretical investigation for five decades. Even for the distributions of the well-studied valence quarks, challenges such as the value of the down quark to up quark ratio at high fractional momenta remain. Much of the sea of quark-antiquark pairs emerges from the splitting of gluons and is well described by perturbative evolution in quantum chromodynamics. However, experiments confirm that there is a non-perturbative component to the sea that is not well understood and hitherto has been difficult to calculate with ab initio non-perturbative methods. This non-perturbative structure shows up, perhaps most directly, in the flavor dependence of the sea antiquark distributions. While some of the general trends can be reproduced by models, there are features of the data that do not seem to be well described. This article discusses the experimental situation, the status of calculations and models, and the directions where these studies will progress in the near future.  
\end{abstract} 

\maketitle
\tableofcontents
\section{Introduction\label{sec:intro}}
\begin{figure}[tbp]
\centering\includegraphics[width=\columnwidth]{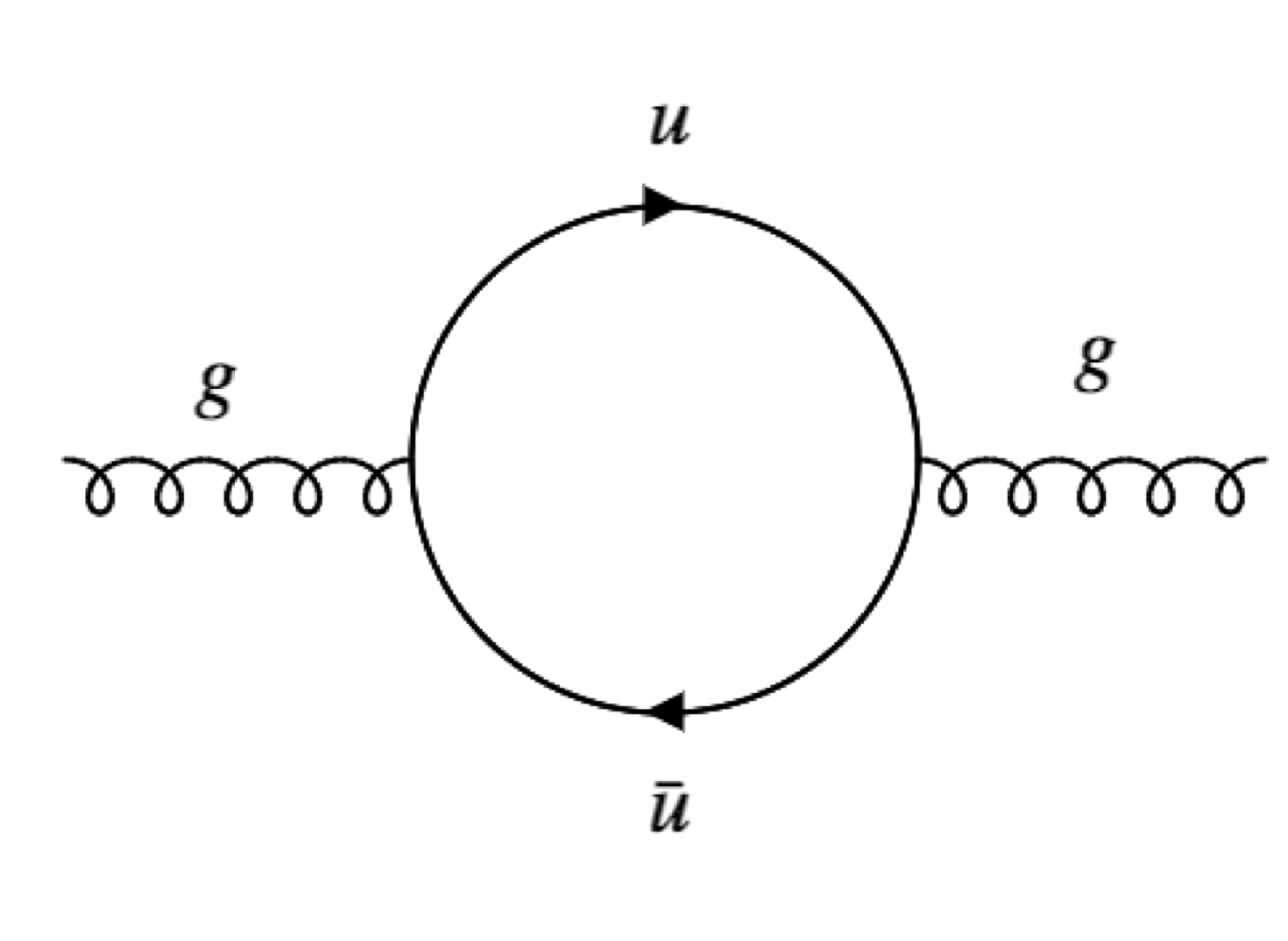} \\[0.5em]
\centering\includegraphics[width=\columnwidth]{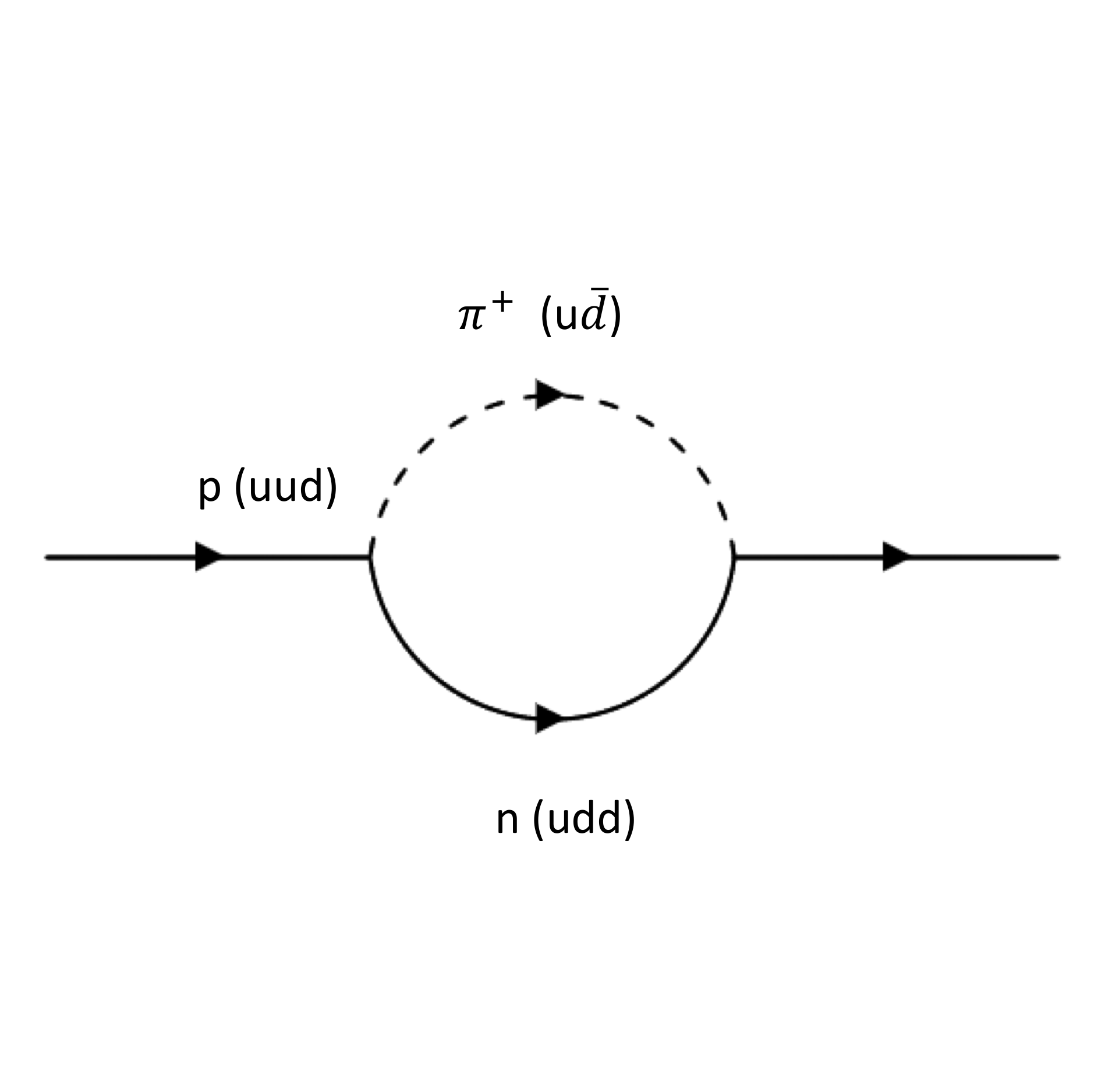}
\caption{ 1a) Sea quarks created in a gluon splitting fluctuation, 1b) Sea quarks created in a pion-nucleon fluctuation
}
\label{fig:1}
\end{figure}

The line in James Joyce’s {\it Finnegans Wake}, ``Three quarks for Muster Mark'', is said to have inspired the choice of name for the three objects proposed to make up a proton and determine its quantum numbers. But experimental data reveal a rich structure of quarks, antiquarks and gluons that are far more abundant than the three so-called valence quarks, especially when the fraction,  denoted as $x$, of the momentum of the proton carried by the quark in a fast moving (infinite momentum) reference frame is small ($x<0.1$). These quarks and anti-quarks are usually referred to as sea quarks (in the early days of the parton model some used the terminology ``ocean'' quarks or ``wee'' partons, perhaps to avoid confusion between sea and charmed $c$ quarks). The presense of antiquarks is natural in quantum field theories as fluctuations of the gluon fields into quark-antiquark pairs as illustrated in Fig.~\ref{fig:1}a. In an even older hadronic picture illustrated in Fig.~\ref{fig:1}b, a proton can fluctuate into, for example, a neutron and a pion, temporarily creating a 5 quark state with two up quarks (u), two down quarks (d) and one down antiquark ($\bar{d}$). In such a non-perturbative picture, the distributions in fractional momentum of sea quarks and sea antiquarks in the proton do not need to have the same shape due to differing masses of the various hadronic components. For any individual u quark, it is not possible to distinguish whether it is a valance or sea quark. The integrals of the distributions of each flavor ( up, down, strange, charm, bottom and top) of quarks over $x$ must obey flavor sum rules for the 2 valence u quarks and one valence d quark in the proton. In a notation that is usually clear in context, ``$\bar{u}$'' can denote a u antiquark, $\bar{u}$,  or the distribution of $\bar{u}$ quarks in fractional momentum, $x$,  $\bar{u}(x)$, where the x dependence has been suppressed. 

For the up quarks
\begin{align}
  \int_0^1 (u(x)-\bar{u}(x))dx = 2 
\end{align}

For the down quarks
\begin{align}
  \int_0^1 (d(x)-\bar{d}(x))dx = 1
\end{align}

For the strange or heavier quarks
\begin{align}
  \int_0^1 (s(x)-\bar{s}(x))dx = 0 
\end{align}

On the other hand the antiquarks and the strange quarks must belong to the sea. Examining their distributions is one of the foci of this review.  If the glue were polarized, the mechanism of Fig.~\ref{fig:1}a transfers that polarization to the sea. On the other hand, in the specific nucleon-pion model of  Fig.~\ref{fig:1}b, the antiquark is contained in a spin-zero object and cannot have an overall  polarization. So a second focus will be on the polarization of the antiquarks.

In the parton model, it was assumed for many years that the mechanism of Fig.~\ref{fig:1}a and iterations of this mechanism dominate the creation of the sea. This assumption is a good approximation at low x and high energy scales. Since the gluons do not carry flavor, the quark-antiquark pair is flavor neutral. One of the successes of quantum chromodynamics (QCD), the theory that describes the interaction of colored quarks and gluons, is that the change in parton distributions with respect to the scale, $\mu^2$, of the interaction, the QCD evolution  (a more precise definition of the meaning of the scale and the evolution equations will be given below), can be quantitatively described. There was speculation that in a proton the Pauli exclusion principle would limit the phase space for the majority quark flavor and lead to more down antiquarks ($\bar{d}$) than up antiquarks ($\bar {u}$), $\bar{d}(x) > \bar{u}(x)$~\cite{Field:1976ve}.  Ross and Sachrajda~\cite{Ross:1978xk} showed that the perturbative QCD evolution contribution to the integral of $\bar{d}(x) - \bar{u}(x)$ was numerically very small and argued that the Pauli blocking was not important (but see Gluck and Reya~\cite{Gluck:2000ch} and the discussion below). Global fits of the distribuions of partons to all the available data built the assumption $\bar{d}(x) = \bar{u}(x)$ into their analyses up to 1989. Gluck and Reya~\cite{Gluck:1977ah} and others, for example~\cite{Parisi:1976fz}, took this a step further and assumed all the sea quarks and glue were generated by QCD evolution, i.e. that at some low scale, $\mu^2_i$,  all the sea quark distributions $(\bar{u}(x,\mu^2_i),\bar{d}(x,\mu^2_i) ,s(x,\mu^2_i) ,\bar{s}(x,\mu^2_i)$ for the up and down antiquarks and the strange and anti-strange quarks) and the gluon distribution $G(x,\mu^2_i)$ were 0. Unless explicitly noted, the contributions for the heavier charm, bottom and top quarks will be ignored. These approaches had considerable success describing the measured parton distributions until the early 1990’s.

The situation changed in 1990 when the New Muon Collaboration (NMC) at CERN first reported~\cite{Amaudruz:1991at} a deep inelastic scattering measurement of the Gottfried sum~\cite{Gottfried:1967} of the difference in the structure functions $F_2$ for the proton and the neutron. With the assumption that the strange quarks contributions are the same in the proton and neutron then
\begin{align}
\int_0^1 \frac{dx}{x} \lf[F_{2}^p(x) - F_{2}^n(x)\rg] = \frac{1}{3} + \frac{2}{3} \int_0^1dx \lf[\bar{u}(x) - \bar{d}(x)\rg].
\end{align}
NMC measured the integral on the left-hand side to be $0.240 \pm 0.016$ showing that $\bar{d}>\bar{u}$. The precise relation between the structure functions $F_2$ and the experimental cross sections for deep inelastic scattering will be given in the next section. Later NMC measurements~\cite{Arneodo:1994sh} led to an updated value of $0.235 \pm 0.026$. It was quickly pointed out in Ref.~\cite{Ellis:1990ti} that the Drell-Yan process~\cite{Drell:1970wh} of hadron-induced di-lepton production would be much more sensitive to the antiquark distributions. The CERN NA51 collaboration measured proton-induced Drell-Yan reactions on targets of hydrogen and deuterium~\cite{Baldit:1994jk} and found in a leading order analysis

\begin{align}
\frac{\bar{u}}{\bar{d}} (\langle x \rangle  = 0.18) = 0.51 \pm 0.04 \pm 0.05.
\end{align}
The NUSEA collaboration at FNAL was able to perform proton-induced Drell-Yan measurements on hydrogen and deuterium over a more extended kinematic range $(0.015 < x < 0.35)$. Their final results for $\bar{d}(x)/\bar{u}(x)$ are shown in Fig.~\ref{fig:2}~\cite{Towell:2001nh} along with the NA51 result. The analysis was based on a next-to-leading order analysis assuming the other parton distributions were well described by contemporaneous global fits(~\cite{Lai:2000a}~\cite{MRST:1998}) and that nuclear corrections for deuterium are small.  Figure ~\ref{fig:3} shows the inferred values of $\bar{d}(x) - \bar{u}(x)$. The resulting integral is 
$\int_{.015}^{0.35} dx \big[\bar{d}(x) - \bar{u}(x)\big] = 0.080 \pm 0.011$ at an average scale of $54\,$GeV$^2$. Extrapolating to the integral from 0 to 1, NUSEA obtained 0.118 $\pm$ 0.012. This value is 10\% of the integrated flavor difference of the valence quarks.  The $x$ dependence of the difference was confirmed by semi-inclusive deep inelastic scattering measurements of the HERMES collaboration~\cite{Ackerstaff:1998} which are also shown in Fig.~\ref{fig:3}.  As will be discussed below, the apparent reduction of the ratio above x of 0.2, admittedly with relatively large uncertainties, is difficult to explain in current models. The behavior of the ratio at larger $x$ will be our third focus.

Since this integral is a flavor non-singlet quantity where the contributions from gluon splitting cancel out in the difference, the result is essentially scale independent. Therefore, there is no scale at which the sea quarks disappear by perturbative evolution, and this flavor difference of the antiquark distributions must be a manifestation of non-perturbative aspects of quantum chromodynamics (QCD). Despite what one may hear, {\bf the proton is never just three valence quarks and glue}.

\section{How to Measure Sea Quark Distributions}
\subsection{Deep Inelastic Scattering}

The relationships between the distributions of the quarks of various flavors and experimental data are covered by essentially all textbooks in high energy and nuclear physics. Here it will be quickly reviewed to define the notation and to point out the salient features of each technique. Figure~\ref{fig:4} illustrates the kinematics for deep inelastic lepton scattering (DIS) with an incident lepton of four momentum $p$ and outgoing momentum $p'$ and a target of four momentum $P$. The momentum transfer through the virtual photon is $q = p - p'$ and $Q^2 = - q^2 > 0$. If the scattering takes place from a very light constituent of mass $m$ carrying a fraction $x$ of the momentum of the target, the squared invariant mass of the quark after the collision is $(xP+q)^2 = x^2P^2 - Q^2 + 2\,x\,P \cdot q \approx m^2 \approx 0$ and the momentum fraction $x = Q^2/(2\,P\cdot q)$. Intuitively (at least to some) if one considers the target in a fast moving reference frame, the lifetime of each virtual state of the target is Lorentz dilated and the longitudinal extent of the target is Lorentz contracted so that a hard (large energy and momentum transfer) probe sees a collection of quarks that is frozen in time with the probability distribution f(x) for each flavor and interacts with the appropriate electro-weak cross section. 

\begin{figure}[t!]
\centering\includegraphics[width=\columnwidth]{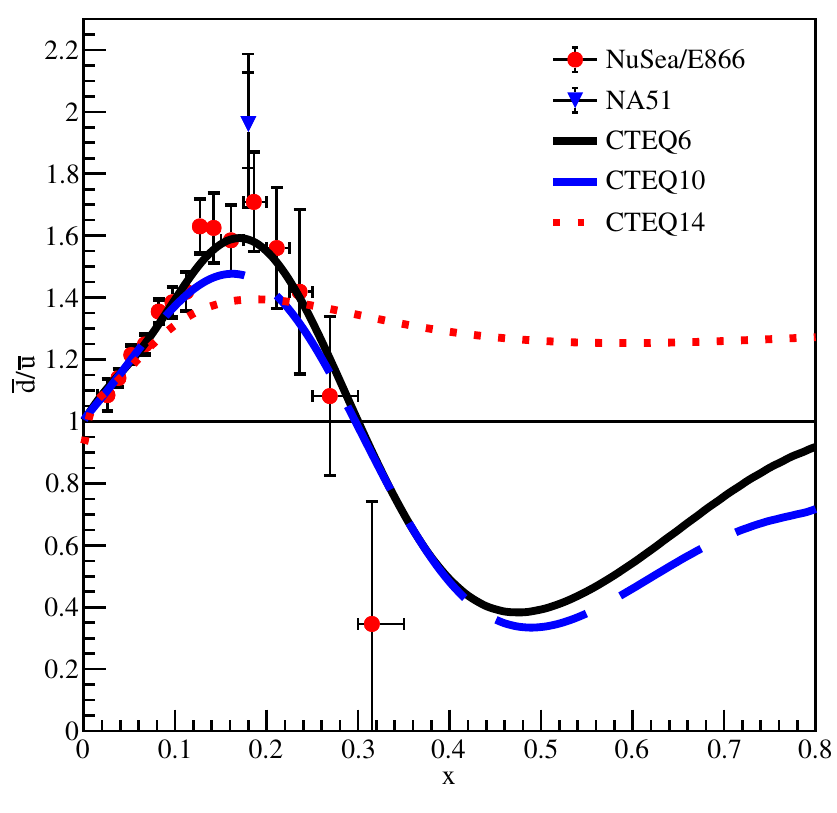}
\caption{The ratios of  $\bar{d}/\bar{u}$ measured by the NUSEA collaboration~\cite{Towell:2001nh} at a scale of 54 GeV$^2$ and NA-51~\cite{Baldit:1994jk} at scales of 25-30 GeV$^2$. The NUSEA analysis was based on a next-to-leading order analysis assuming the other parton distributions were well described by contemporaneous global fits(~\cite{Lai:2000a}~\cite{MRST:1998}) and that nuclear corrections for deuterium are small. The curves are next-to-leading order global fits of CTEQ6, CTEQ10~\cite{Lai:2010vv} and CTEQ14~\cite{Dulat:2015mca} in $\overline{MS}$ renormalization scheme, all at scales of 54 GeV$^2$, to show how the parameterizations have changed over time, especially in the unmeasured region.
}
\label{fig:2}
\end{figure}

\begin{figure}[tbp]
\centering\includegraphics[width=\columnwidth]{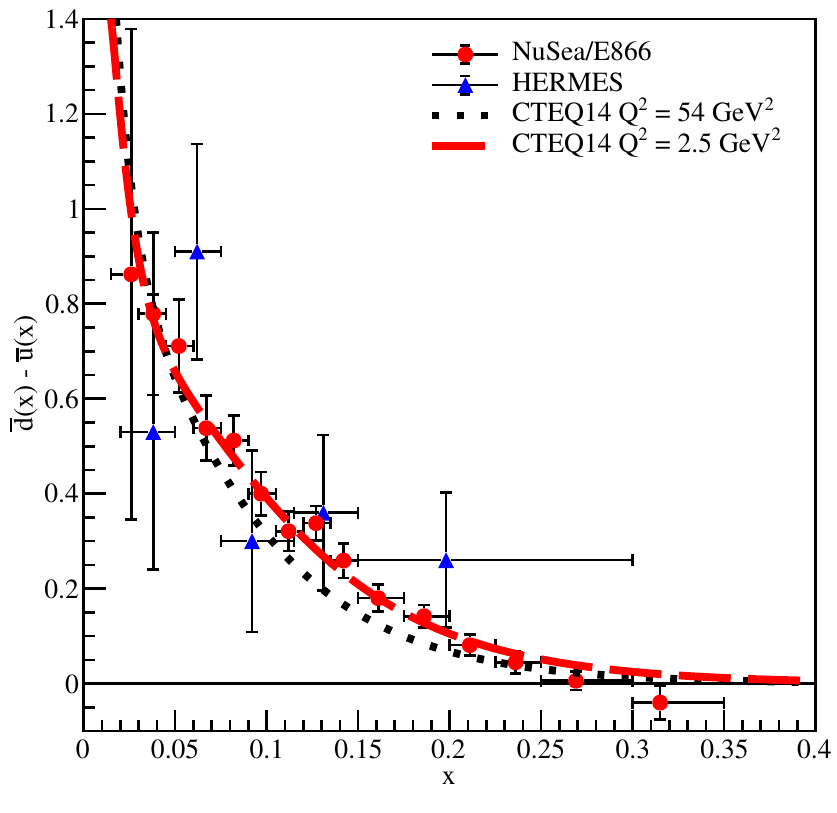}
\caption{Results for $\bar{d}(x)-\bar{u}(x)$ (red circles) at a scale of 54 GeV$^2$ obtained from the NUSEA~\cite{Towell:2001nh} ratio measurements shown in Fig. 2 and CTEQ5M~\cite{Lai:2000a}  or MRST~\cite{MRST:1998} parameterizations of $\bar{d}+\bar{u}$. The blue triangles are the semi-inclusive deep inelastic scattering results from the HERMES collaboration~\cite{Ackerstaff:1998} in leading order at a scale of 2.5 GeV$^2$. Given the size of the error bars, for this qualitative comparison the warning below not to mix results at different scales, renomalization schemes, and order of perturbative expansion in $\alpha_s$ has been ignored.  The CTEQ14~\cite{Dulat:2015mca}              next-to-leading order global fit results are also shown at scales of 54 GeV$^2$ and 2.5 GeV$^2$ to illustrate the size of the scale effect.
}
\label{fig:3}
\end{figure}

It can be proven for deep inelastic scattering that the cross section factorizes. (For details, see the discussion for example in Ref.~\cite{Brock:1993sz}. )
\begin{align}
\s^h = \sum_{i=f,\bar{f},G} \int_0^1 \frac{d\xi}{\xi}\,
C^i\lf(\frac{x}{\xi},\frac{Q^2}{\mu^2},\frac{\mu_f^2}{\mu^2},\a_s(\mu^2)\rg) \ \phi_{i/h}\lf(\xi,\mu_f,\mu^2\rg)
\end{align}
where the sum is over all quark flavors and glue. The $C^i$ are hard scattering functions that are ultraviolet and infrared safe and calculable in perturbation theory. They are a function of quark flavor, the physical process (for example the nature of the vector boson being exchanged in DIS and the order of perturbation theory of the calculation), the renormalization scale $\mu^2$, the factorization scale $\mu_f^2$ and the strong coupling constant $\a_s$, but not the distribution of partons. The renormalization scheme eliminates the ultraviolet divergences of the hard scattering amplitude. The parton distributions for each flavor i, $\phi_{i/h}$, contain all the infrared sensitivity, are specific to the particular hadron, $h$, and depend on the factorization scale $\mu_f$ and the factorization scheme, but do not depend on the hard scattering process. If defined consistently, they are universal. Therefore, one can combine data from different kinds of experiments to determine the parton distributions. The factorization scale defines the separation of short-distance and long-distance effects.  By convention, one often sets the renormalization and factorization scales in deep inelastic scattering to $Q^2$, but that is not necessary. For hadron-hadron reactions, one must integrate over the parton distributions of both the beam and target, but the separation of the hard scattering functions from the parton distributions remains. In such reactions, the choice of renormalization and factorization scales to be used is less obvious. In all cases, since the hard scattering functions depend on the order of perturbation theory, the scheme and scales, the parton distributions  are not directly physical observables. It is not consistent to use parton distributions obtained from, for example, fits using next to leading order hard scattering functions in calculations done at a different order or to directly compare various parton distribution functions when they are not defined consistently.

\begin{figure}[tbp]
\centering\includegraphics[width=\columnwidth]{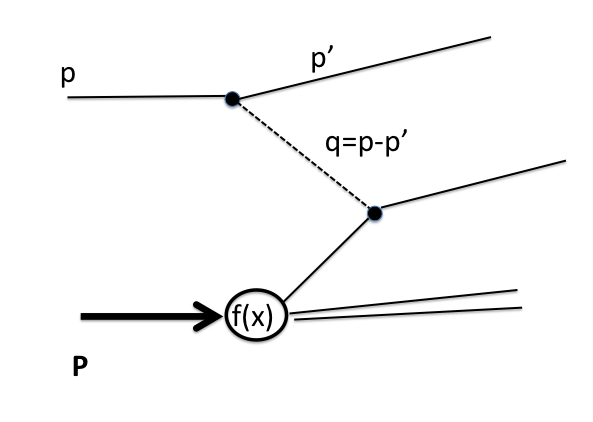}
\caption{Deep inelastic scattering Feynman diagram and kinematic variables.
}
\label{fig:4}
\end{figure}

Another important feature of QCD is that if the factorization and renormalization scales are taken as $\mu^2 = \mu_f^2 = Q^2$, then QCD allows one to calculate the parton distributions at higher $Q^2$. This is the DGLAP QCD evolution of Dokshitzer~\cite{Dokshitzer:1977sg}, Gribov, Lipatov~\cite{Gribov:1972ri}, Altarelli and Parisi~\cite{Altarelli:1977zs}.
\begin{multline}
\mu \,\frac{d\,\phi_{i/h}\lf(x,\mu_f,\mu^2\rg)}{d\mu} = \\
\sum_{j=f,\bar{f},G} \int_x^1\frac{d\xi}{\xi}\,
P_{ij}\lf(\frac{x}{\xi},\a_s(\mu^2)\rg)\,\phi_{i/h}\lf(\xi,\mu_f,\mu^2\rg).
\end{multline}
The explicit forms of the splitting functions, $P_{ij}$, which describe gluon emission and absorption from the quarks, gluon splitting to two quarks and recombination, and gluon-gluon interactions can be found in many textbooks and references (for example Ref.~\cite{Brock:1993sz}). While this is an important test of QCD, it also allows data taken at different $Q^2$ to be usefully combined to determine parton distributions. Moreover, the excellent quantitative agreement with the data indicate
that Pauli blocking effects are not important at large $Q^2$ since Pauli blocking is not included in the DGLAP equations. 

At extremely small x, the DGLAP evolution loses its validity, and it needs to be combined with the BFKL resummation of small-x logarithms to all orders of perturbation theory. Recent references~\cite{Ball:2018a}~\cite{Abdolmaleki:2018a} show the need for this extension in the low x HERA data and explore the impact for the LHC.  

The cross sections for neutral-current electron and muon deep inelastic scattering  (where the small Z exchange contribution is ignored) and neutral- and charged-current neutrino deep inelastic scattering from an unpolarized target are
\begin{align}
\frac{d^2\,\s_i}{dx\,dy} &= \frac{4\,\pi\,\a^2}{x\,y\,Q^2}\,\eta^i
\Bigg\{\lf(1 - y - \frac{x^2 y^2 M^2}{Q^2}\rg)F_{2}^i(x) \no \\
&\hs*{20mm}
+ x\,y^2\,F_1^i(x) \mp \lf(y - \frac{y^2}{2}\rg)x\,F_3^i(x)\Bigg\},
\end{align}
where $i$ labels either neutral current (NC) or charged current (CC) and the exchanged vector boson, $\g,\,Z$ or $W^\pm$. $y = \frac{p\cdot q}{P \cdot q} = \frac{v}{E}$ is the fraction of the lepton’s energy loss in the target rest frame, and $\a$ is the fine structure constant. The sign of the last term, which is parity violating, is $-$ for antineutrinos and $+$ for neutrinos. $\eta^i$ is the relative coupling strength for the weak interaction compared to the electromagnetic interaction.
\begin{align}
\eta_\g^{NC} &= 1, \\
\eta_Z^{NC} &= \lf(\frac{G_F\,M_Z^2}{2\sqrt{2}\,\pi\a}\rg)^2\lf(\frac{Q^2}{Q^2 + M_Z^2}\rg)^2, \\
\eta_W^{CC} &= 4\lf(\frac{G_F\,M_W^2}{4\,\pi\,\a}\,\frac{Q^2}{Q^2 + M_W^2}\rg)^2,
\end{align}
where $G_F$ is the Fermi constant, $M_Z$ and $M_W$ the masses of the neutral and charged weak intermediate vector bosons.

At lowest order in terms of the partons for deep inelastic scattering, the $C^i$ are just determined by the electroweak interaction, and with the approximation of the Callan-Gross relation that $F_2 = 2x\,F_1$ then
\begin{align}
F_2^\g &= x\sum_q\,e_q^2\lf[q(x) + \bar{q}(x)\rg],
\end{align}
where $e_i$ are the appropriate electric charges of each quark flavor.  For incident $\bar{\nu}$ on a proton
\begin{align}
F_2^{W^-} &= 2\,x\lf[u(x) + \bar{d}(x) + \bar{s}(x) + c(x)\rg],\\
F_3^{W^-} &= \ph{x\,}2\lf[u(x) - \bar{d}(x) - \bar{s}(x) + c(x)\rg].
\end{align}
In these expressions, weak quark flavor mixing (Cabbibo-Kobayashi-Maskawa mixing) and quark mass threshold effects  have been ignored. To obtain the structure functions for an incident $\nu$, interchange $d$ with $u$ and $s$ with $c$ and similarly interchange the antiquark flavors. Based on charge symmetry, that the masses of the u and d quarks are much lighter than any other scale in the proton, interchanging $u \leftrightarrow d$ and $\bar{u} \leftrightarrow \bar{d}$ distributions gives the structure functions for scattering from a neutron.

In principle, the parity violating term $F_3$ for charged current neutrino scattering on an isoscalar target gives a good measurement of the valence distribution while the  $F_2$ term from either neutral or changed current interactions measures the sum of valence plus sea. In practice, the valence distributions are larger than the sea distributions for $x>0.04$ which magnifies the errors in determining the sea in this way. A more substantive issue is that most of the high statistics neutrino experiments are performed with heavy targets such as iron. One must deal with the nuclear corrections that, despite 30 years of study, are still not well understood. Nuclear corrections must also be considered for the ``neutron'' data that are taken from results on a deuterium target, often with the assumption that the nuclear effects are small. Recent work relating short-range correlations to the size of the nuclear effects has led some to predict that the nuclear effects in deuterium are larger than previously believed at higher $x$ values~\cite{Weinstein:2010rt}.

\begin{figure}[tbp]
\centering\includegraphics[width=\columnwidth]{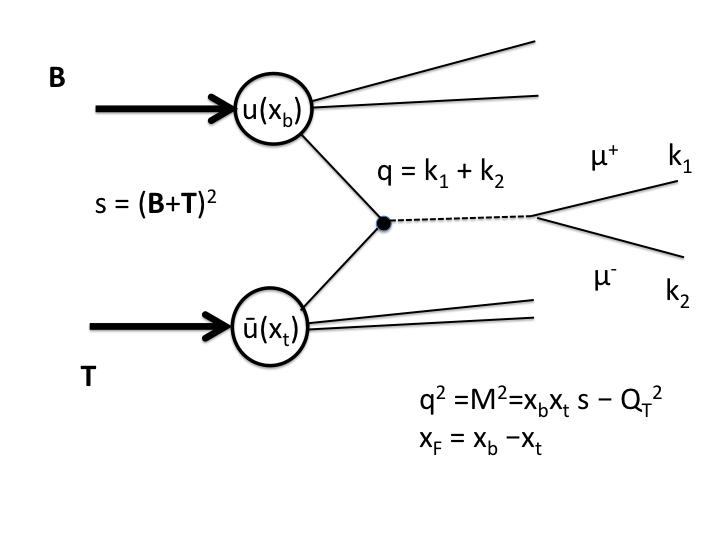}
\caption{Drell-Yan Feynman diagram and kinematic variables.
}
\label{fig:5}
\end{figure}

\subsection{Drell-Yan Reactions}
A much more direct method to study the sea of antiquarks is to use a hadron-induced reaction and detect a virtual photon. This is the diagram shown in Fig.~\ref{fig:5} for the Drell-Yan process where a quark from the beam annihilates an antiquark in the target or vice versa. Factorization theorems also have been proven for this reaction~\cite{Bodwin:1984hc}. The leading order cross section can simply be written as
\begin{align}
\frac{d\s}{dx_b\,dx_t} = \frac{8\,\pi\,\a^2}{9\,s\,x_b\,x_t}\,\sum_q\,e_q^2
\lf[q(x_b)\bar{q}(x_t) + \bar{q}(x_b)q(x_t)\rg].
\end{align}
$x_b$ and $x_t$ are the momentum fractions of the beam and target partons participating in the reaction, and $s$ is the square of the center of mass energy of the beam and target. If an experiment is performed, for example, with a proton beam that selects $x_b$ in the valence dominated region and $x_F = x_b - x_t \gg 0$, the first term dominates and the charge squared weighting and the fact the $u_v(x) \sim 2\,d_v(x)$ means the measurement is, by a factor of $\sim 8$, more sensitive to $\bar{u}$ quarks in the target than $\bar{d}$. The scale is usually chosen as the mass squared of the virtual photon, $M^2 = x_b\,x_t\,s - Q_T^2$ where $Q_T^2$ is the square of the transverse momentum of the virtual photon and is usually small compared to $M^2$. Again, using charge symmetry ($\bar{u}_p = \bar{d}_n$ , $\bar{d}_p = \bar{u}_n$ and assuming the nuclear corrections in the deuteron are small, the ratio of the cross section on deuterium to that on hydrogen directly measures $\bar{d}(x)/\bar{u}(x)$.
\begin{align}
\frac{\s_d}{\s_p} \approx \frac{\s_p + \s_n}{\s_p} \approx 1 + \frac{\bar{d}_p}{\bar{u}_p}.
\end{align}
Calculations of the effect of nuclear corrections for deuterium~\cite{KamLee:2012}~\cite{Ehlers:2014} show they are small for the x range currently measured. It is known that the next-to-leading order QCD corrections to the Drell-Yan cross section are substantial, approximately a factor of two (emphasizing the need for consistent use of hard scattering functions and parton distributions to the same order in $\a_s$), but the corrections factors for the proton and neutron are very similar. The NUSEA results in Fig.~\ref{fig:2} are based on a next-to-leading order calculation involving all quark flavors, but the results are very close to what are obtained from the simple leading order formulae discussed here as long as the actual $x_F$ distributions of the data are considered (i.e. whether $x_F \gg 0$. For example, the NA-51 data have $<x_F> \approx 0.$ ).

\subsection{Global Fits}

\begin{figure}[tbp]
\centering\includegraphics[width=\columnwidth]{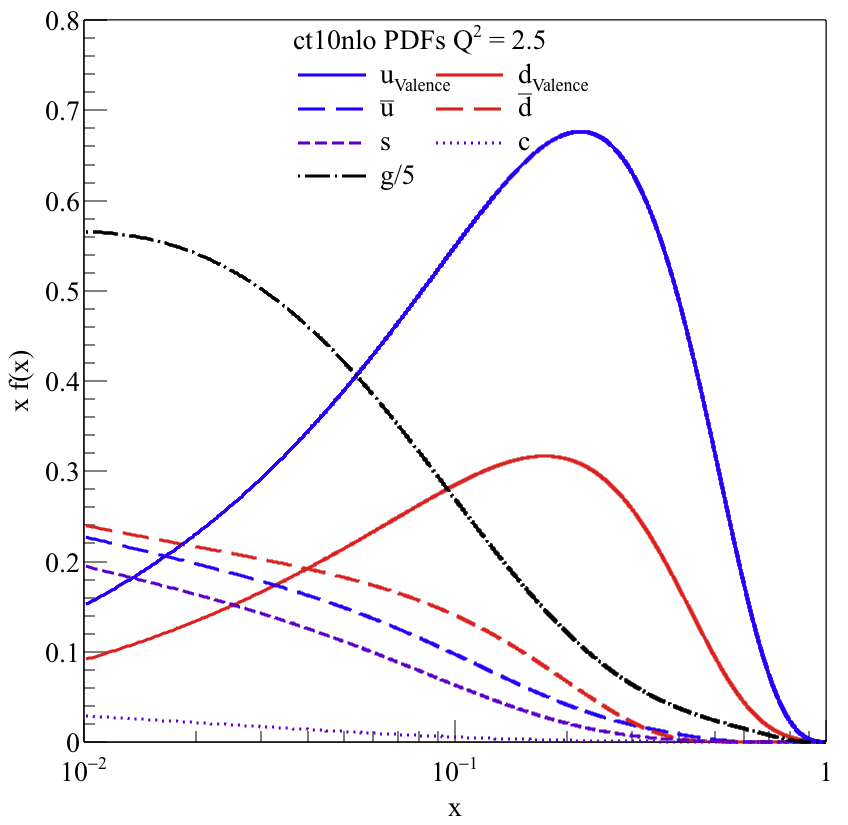} \\[0.5em]
\centering\includegraphics[width=\columnwidth]{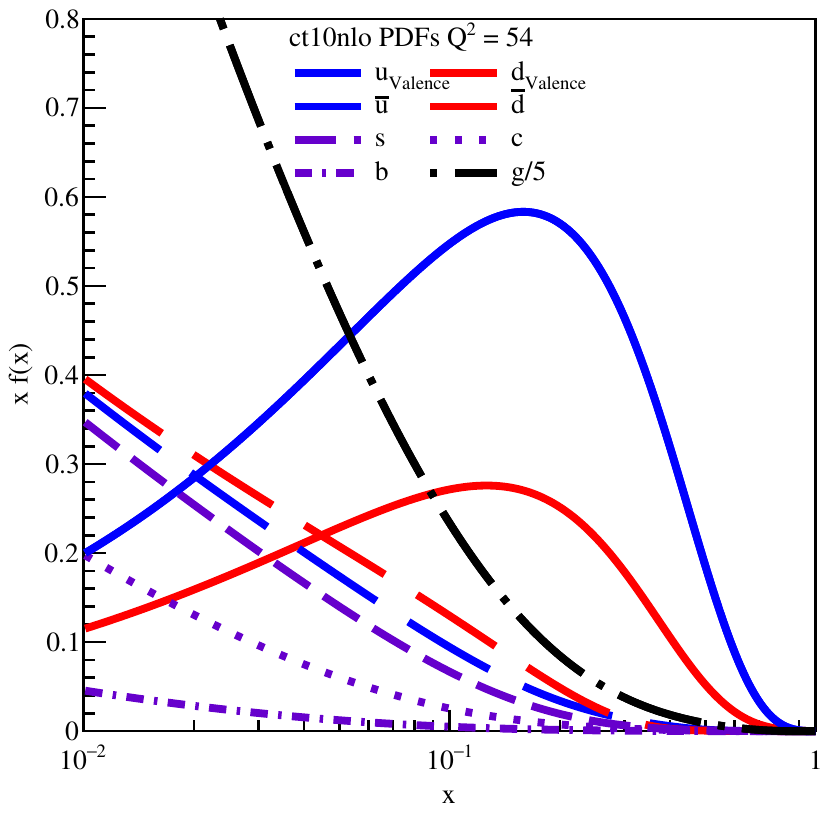}
\caption{CTEQ10 next-to-leading order parton distribution functions~\cite{Lai:2010vv} at upper) $Q^2$ = 2.5 GeV$^2$ and lower) 54 GeV$^2$. Note that the gluon distributions have been divided by a factor of 5.
}
\label{fig:6}
\end{figure}

In major efforts, several groups have performed extensive systematic fits of parton distribution functions (pdf) to all the available deep inelastic scattering and Drell-Yan data and report error bars on the results. These include the Coordinated Theoretical-Experimental Project on QCD (CTEQ); Martin, Roberts, Sterling and Thorne (MRST); and the Neural Networks PDF collaboration (NNPDF)~(next-to-leading order ~\cite{Ball:2008a}~\cite{Ball:2010a} and next-to-next to leading order~\cite{Ball:2015a}). Analyses of $W$ and $Z$ boson production at the LHC discussed below are influenced by the next-to-next to leading order corrections. The global fits are periodically updated as new data become available, so many versions exist, usually with distinct labels such as CTEQ6~\cite{Pumplin:2002vw}, CTEQ10~\cite{Lai:2010vv} and CTEQ14~\cite{Dulat:2015mca}. One challenge for all but the Neural Networks PDF collaboration is understanding the correlations inherent in their assumptions of a functional form to fit. The other challenge is incorporating the systematic errors of the various data sets properly. There are often tensions between the various data sets that are pointed out in the articles reporting the fits, and options are explored emphasizing or deemphasizing one or another of the experimental results. When used in QCD calculations, these global fits are very successful in describing a wide variety of data from collider and fixed target experiments. Examples of the parton distributions extracted by CTEQ10~\cite{Lai:2010vv} at two different scales are shown in Fig.~\ref{fig:6}.  The integrals of the flavor differences of the sea contributions from various global fits and models are given in Table 1. Extending the integral down to $x$ of 0 does require some care, so in Tab. 1 it is generally cut off at some lower $x$ value. For example, in the CTEQ14 fit, $\bar{d}-\bar{u}$ changes sign at $x \sim 0.006$. The integral in the CTEQ10 fit is relatively stable as the lower x limit is pushed down to, for example, 0.00001.  
\begin{table}
\begin{tabular}{|cccccc|} \hline
  $x_{min}$ & $x_{max}$ & $\int_{x_{min}}^{x_{max}}(\bar{d}-\bar{u})dx $& Q$^2$  &Source & Ref. \\
                    &.                 &                                                                         &(GeV$^2$) &  &\\
    0.0 & 1.0 & 0.147 $\pm$ .026 & 4 &NMC & \cite{Arneodo:1994sh} \\
    0.015 & 0.35 & 0.080 $\pm$ 0.011 & 54 & NUSEA & \cite{Towell:2001nh} \\
    0.0 & 1.0 & 0.118 $\pm$ 0.012 & 54 & NUSEA & \cite{Towell:2001nh} \\
    0.001 & 1.0 & 0.165           & 54 & CT66nlo & \cite{Pumplin:2002vw} \\
    0.001 & 1.0 & 0.114         & 54 & CT10nlo & \cite{Lai:2010vv} \\
    0.001 & 1.0 & 0.116         & 2  & CT10nlo &  \cite{Lai:2010vv} \\
    0.01 & 1.0 &  0.090         & 54 & CT14nlo & \cite{Dulat:2015mca} \\
    0.001 & 1.0 & 0.086 & 1  & Stat. Mod. & \cite{Bourrely:2015kla} \\
    0. &  1.0   & 0.13  & ?  &Det. Bal. & \cite{Zhang:2010prd} \\
    0.02 & 0.345 & 0.108 & 54 & Chiral Soliton & \cite{Pobylitsa:1999} \\
    0.0 & 1.0 & 0.13 $\pm$ 0.07 & ?  &Lattice &\cite{Lin:2016}\\  \hline
\end{tabular}
\caption{Integrals of $(\bar{d}-\bar{u})$ from $x_{min}$ to $x_{max}$ from experiment (NMC and NUSEA) and from  several global fits (CTEQ6.6, CTEQ10, CTEQ14), calculations (Lattice), and models (Statistical and Detailed Balance). The weak variation of the integral to the choice of scale is illustrated with the CTEQ10 comparison at 2 and 54 GeV$^2$. The scales of the detailed balance and lattice calculations are not explicitly reported in those references.}
\end{table}

\subsection{Spin-Dependent Parton Distributions}

With polarized beam and target, additional spin-dependent parton distributions are needed to fully characterize the nucleon response.  With a longitudinal polarized lepton beam incident on a longitudinally polarized nucleon target, the asymmetry of spin-antiparallel cross sections $(\s^{1/2})$ to spin-parallel cross sections $(\s^{3/2})$ divided by the sum is

\begin{align}
  A(x,Q^2) = \frac{ d^2 \s^{1/2}-d^2 \s^{3/2}}{ d^2 \s^{1/2} + d^2 \s^{3/2}} = D [ A_1+\eta A_2] \approx D \frac{g_1(x,Q^2)}{F_1(x,Q^2)}, 
\end{align}

\begin{align}
  D = \frac{2y-y^2}{2(1-y)(1+R)+y^2},
\end{align}

\begin{align}
  \eta = \frac{\sqrt{Q^2} 2 (1-y)}{E y (2-y)},
\end{align}

\begin{align}
  g_1(x,Q^2) = \sum_i e_i^2 ( q^{\uparrow}(x,Q^2)-q_i^{\downarrow}(x,Q^2)) = \sum_i e_i^2 \Delta q_i(x,Q^2),
\end{align}
where the $q^{\uparrow / \downarrow}(x,Q^2)$ are the quark helicity distributions. E is the incident beam energy (for an experiment with the target at rest in the lab), D is the virtual photon polarization, and R is the ratio of the longitudinal to transverse cross section. $A_2$ is bounded by R which is small in the Callan-Gross limit and $\eta$ is usually small, so $A_2$ usually gives a relatively small contribution to $A(x,Q^2)$. The global analyses can then be extended to polarized structure functions, though the polarized data are much sparser. (See for example Ref. \cite{NNPDF:pol1-1}.)

  While not directly measurable in inclusive polarized lepton scattering on a polarized target, the third twist-two parton distribution can be accessed in semi-inclusive deep inelastic scattering or with transversely polarized hadron beams on transversely polarized nucleon targets. These are known as the transversity distributions, $\delta q(x,Q^2)$.  So far little data on the tranversity distributions are available. Several groups have been able to extract limited valence transversity distributions~\cite{Anselmino:2013a}~\cite{Bacchetta:2013}~\cite{Martin:2015}. Only Martin, Bradamante and Barone~\cite{Martin:2015} have obtained sea quark transversity distributions. They are typically smaller than the valence distributions and within the current error bars, consistent with 0. In a non-relativistic model,  $\delta q(x,Q^2)$ can be easily obtained by rotating $ \Delta q_i(x,Q^2)$, but in a relativistic treatment of rotations this is no longer true. 

\subsection{Parameterizing Parton Distributions}

Historically, the evolution of thinking about the sea generally took the following path. At high $x$, the valence quarks dominate, and there was little experimental information about the sea. Based on the models of Regge theory at low x and quark counting rules at high $x$, for the valence quarks $x\,f(x)$ was expected to be proportional to $ x^{1/2}\,(1-x)^{2n-1}$ where the number of spectator quarks, $n$, equals $2$. A typical functional form for the global parton distribution fitting was $x\,f(x) = C x^\a(1-x)^\b$ with $C$, $\a$, and $\b$ as free parameters for each flavor. (Modern fits of parton distributions find more general parameterizations are required for high quality reproduction of the body of experimental data.) Based on the approximately constant photon-proton total cross sections at high energy, for the sea $x\,f(x)$ is approximately constant at low $x$ and $Q^2$. Since the existence of sea quarks imply at least a 5 quark Fock state, the $x\,f(x)$ at high $x$ were expected to behave as $(1-x)^{2 \times 4-1}$, falling off rapidly at high $x$ and being almost $x$ independent at low $x$. Again a similar functional form was assumed. A recent global analysis confirms these expectations for nucleon parton distributions at low x and for the valence quarks at high x, but for the sea and glue at high x the agreement is only qualitative~\cite{Ball:2016a}. It should be noted that there is still debate about the quark counting rules for the pion parton distribution functions at high x and soft gluon summation seems to be important~\cite{Wij:2005}~\cite{Aicher:2010}.

On an isoscalar target
\begin{align}
\frac{F_2^{\nu p} + F_2^{\nu n}}{2} &= x\lf[u(x) + \bar{u}(x) + d(x) + \bar{d}(x) + s(x) + \bar{s}(x)\rg], \\
\frac{F_2^{e p} + F_2^{e n}}{2} &= \no \\
&\hs*{-15mm}
\frac{5}{18}\,x\lf[u(x) + \bar{u}(x) + d(x) + \bar{d}(x) + \frac{2}{5}\lf[s(x) + \bar{s}(x)\rg]\rg].
\end{align}
IF the contributions of the strange quarks were small, the average of the $F_2$'s for neutrinos would be $18/5$ times the $F_2 $'s for electromagnetic processes. When the data showed this approximate relation held for $x>0.1$, it was concluded that the assumption of small strange quark contributions was valid.

At low x, the experiments at HERA found a rising cross section showing that the glue must dominate for $x$ below $\sim$0.01 and high $Q^2$ as shown in Fig. 6.  In that case, gluon splitting dominates the antiquark distributions and for a given flavor
\begin{align}
  \bar{u}(x,Q^2) \approx \bar{d}(x,Q^2) \approx s(x,Q^2) \approx \bar{s}(x,Q^2).
\end{align}
\begin{eqnarray} 
 \bar{u}(x,Q^2)  &\approx & \frac{\a_s}{2 \pi} \int_x^1 \frac{dy}{y} \nonumber \\
   &  &    \left[ G(y) P_{qg}(\frac{x}{y}) + \bar{u}(y)(\delta(1- \frac{x}{y})+ P_{qq}(\frac{x}{y})) \right]\ log \frac{Q^2}{\mu^2} \nonumber \\
   &. &
\end{eqnarray}
where $G(y)$ is the gluon distribution. The splitting function $P_{qg} (z)$ is the probability a gluon annihilates into a $q \bar{q}$ pair where the quark has a fraction, z, of the momentum of the gluon. $P_{qq} (z)$ is the probability a quark splits into a quark of momentum fraction z  and a gluon.

\begin{align}
  P_{qg}(z) = \frac{1}{2}(z^2 + (1-z)^2)
\end{align}
\begin{align}
  P_{qq}(z) = \frac{4}{3} \frac{1+z^2}{1-z}
\end{align}

\subsection{Strange Quark Distributions}

Eq. 23 assumes that at low $x$ (high energy) the strange quark mass is negligible.
Information on the strange sea at low $x$ values comes from $W$ production at $p+p$ colliders, LHC and RHIC. The formalism is exactly parallel to that of Drell-Yan production with the experimental difference that the neutrino from the leptonic $W$ decay branches is not detected but inferred from missing energy. The results from ATLAS~\cite{Aaboud:2016btc} are also included in Fig.~\ref{fig:9} below at their x value of maximum sensitivity, $\sim \, $0.023 at a scale of 1.9 GeV$^2$ 
\begin{align}
\frac{s + \bar{s}}{2 \bar{d}} &= 1.19 \pm 0.07\,(\text{exp.}) \pm 0.02\,(\text{mod}){}^{+0.02}_{-0.10}\,(\text{par}),
\end{align}
where (mod) indicate uncertainties from model variation and (par) are from parameter variations. It indicates at such $x$ values the effects of quark mass on gluon splitting are indeed small.

To gain better sensitivity to the strange quarks, one can look for neutrino DIS events with two opposite sign muons in the final state. These were expected to arise mainly from the processes
\begin{align}
\nu + s \to \mu^- + c \quad \text{and then} \quad c \to \mu^+ + \nu_\mu + s, \\
\bar{\nu} + \bar{s} \to \mu^+ + \bar{c} \quad \text{and then} \quad \bar{c} \to \mu^- + \bar{\nu}_\mu + \bar{s}.
\end{align}
Often it was assumed the strange quark distributions at higher x had the same shape in $x$ as $(\bar{u} + \bar{d})$, differing only by a scale factor.
In a leading order analysis by the NuTeV collaboration~\cite{Goncharov:2001qe}, the strange sea parton distributions  were assumed to be
\begin{align}
  s(x) =  \kappa_{\nu} (1-x)^{\alpha_{\nu}} \left [ \frac {\bar{u}(x)+\bar{d}(x)}{2} \right] 
\end{align}
\begin{align}
  \bar{s}(x) = \kappa_{\nu}^- (1-x)^{\alpha^-_{\nu}} \left[ \frac {\bar{u}(x)+\bar{d}(x)}{2} \right] 
\end{align}

$\kappa$=1 and $\alpha_{\nu}$ = 0 would correspond to a SU(3) flavor symmetric sea. Typical fitted results for $ \kappa_{\nu}$ and $\kappa_{\nu}^-$ are 0.44 $\pm$ 0.06 $\pm$ 0.04 and 0.45 $\pm$ 0.08 $\pm$ 0.07, respectively.
The results are sensitive to the choice of the non-strange parton distributions. For example, with GRV~\cite{Gluck:1994uf} non-strange distributions, $ \kappa_{\nu}$ and $\kappa_{\nu}^-$ are 0.37 $\pm$ 0.05 $\pm$ 0.03 and 0.37 $\pm$ 0.06 $\pm$ 0.06.

\begin{figure}[tbp]
\centering\includegraphics[width=\columnwidth]{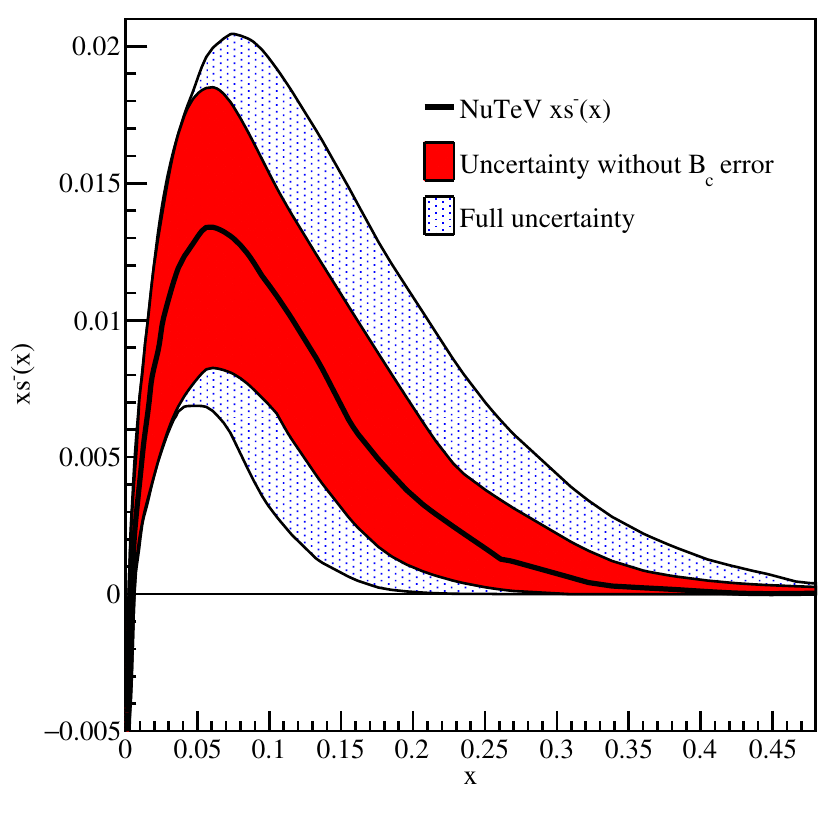}
\caption{The results of NuTeV~\cite{Mason:2007zz} fits for $xs^-(x)=x(s(x)-\bar{s}(x)$ at $Q^2$ of 16 GeV$^2$. The inner band is the uncertainly without including that of the charm semi-leptonic branching ratio. The outer band is the combined error.  
}
\label{fig:7}
\end{figure}

Thus, the higher $x$ measurements of the strange sea, which chronologically came first, revealed a different picture, showing that the strange quarks are suppressed relative to $\bar{u}$ and $\bar{d}$. Again, this is usually interpreted as due to the influence of the heavier strange quark mass at lower energies. Hadronic models would suggest that $s(x) \neq \bar{s}(x)$ since the mass of the $uds + u\bar{s}$ fluctuation $(\Lambda + K^+)$ is lower than that of any system with an $\bar{s}$ quark in the baryon; thus they would have different $x$ dependences. In a next-to-leading order analysis ,the NuTeV collaboration~\cite{Mason:2007zz} found
\begin{align}
\int_0^1 dx\, x\lf[s(x) - \bar{s}(x)\rg] &= 
0.00196 \pm 0.00046\,(\text{stat.}) \no \\
&\hs{-8mm}
\pm 0.00045\,(\text{syst.}){}^{+0.00148}_{-0.00107}\, (\text{external}).
\end{align}
The external error refers to uncertainties on external measurements such as the charm quark mass and the charm semi-leptonic branching ratio. The fitted $x$ dependence of this analysis is shown in Fig.~\ref{fig:7}. The $x$ dependence of the uncertainty band is partially a consequence of the assumed functional form. 

\begin{figure}[tbp]
\centering\includegraphics[width=\columnwidth]{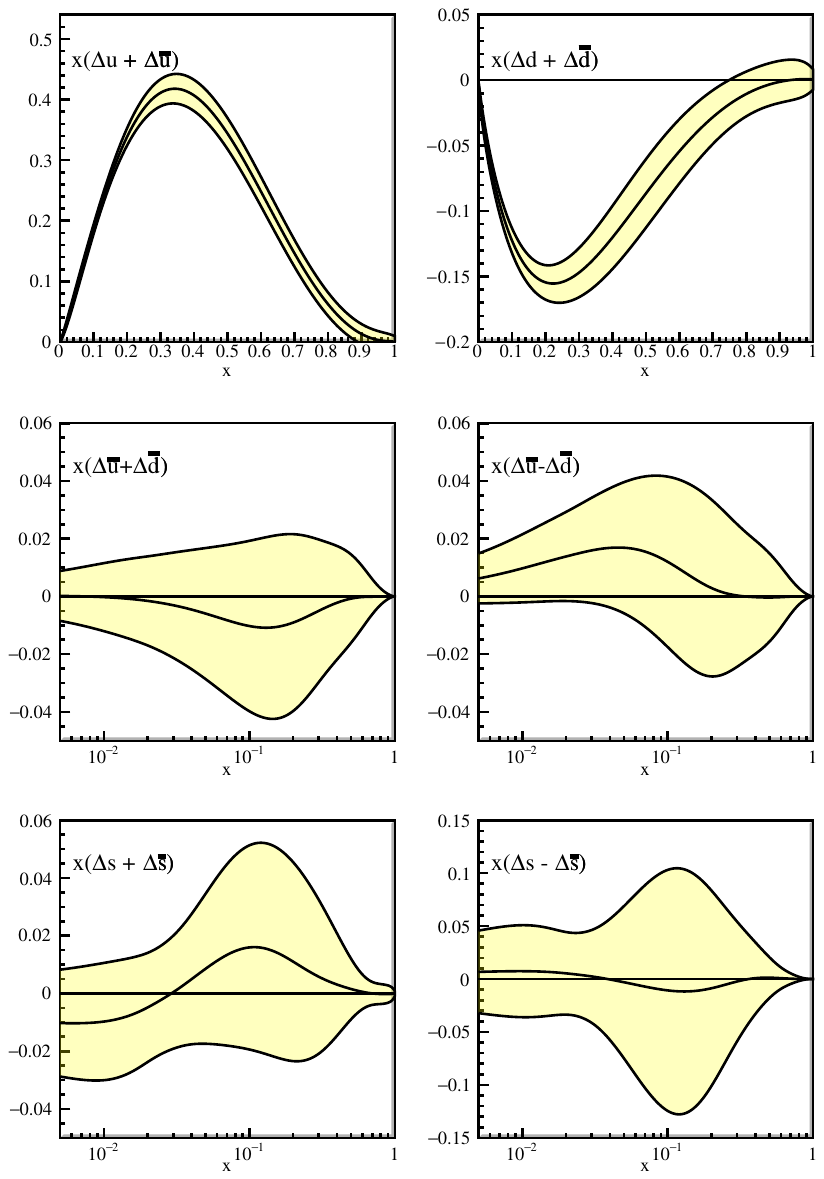}
\caption{Spin-dependent parton distribution functions with 1 $\sigma$ uncertainty bands at a scale of 1 GeV$^2$ of Ethier, Sato and Melnitchouk ~\cite{Ethier:2017zbq} obtained from a self-consistent fit of parton distributions and fragmentation functions.   
}
\label{fig:8}
\end{figure}

\begin{figure}[tbp]
\centering\includegraphics[width=\columnwidth]{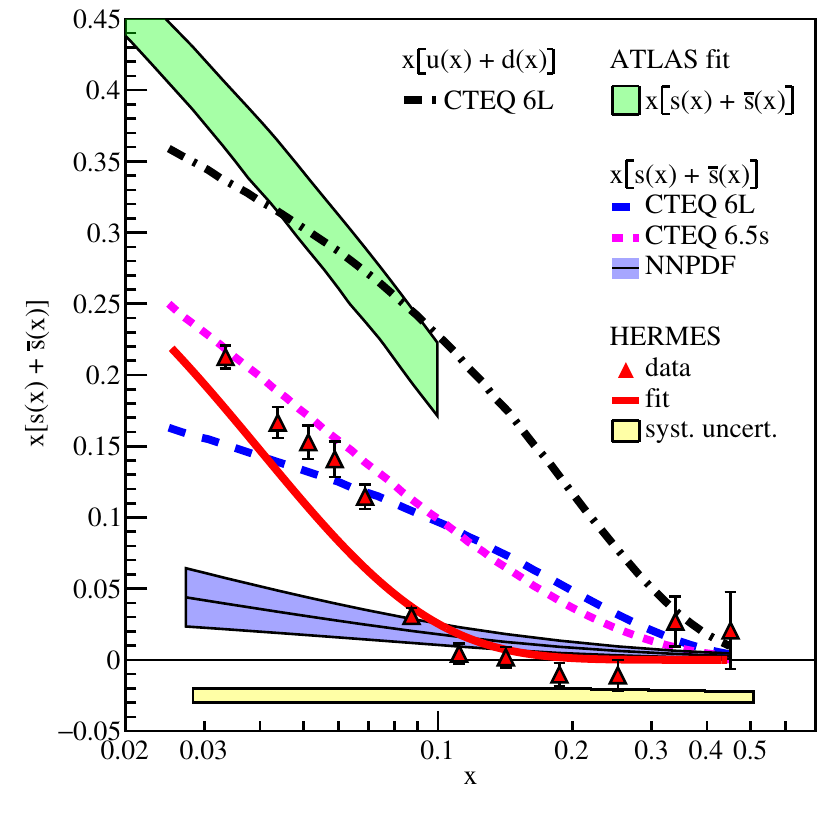}
\caption{ $x(s(x)+\bar{s}(x))$ obtained by HERMES~\cite{Airapetian:2013zaw} from a leading order analysis of semi-inclusive kaon production on deuterium at a scale of 2.5 GeV$^2$. The dotted black and blue lines are the CTEQ6L fits~\cite{Pumplin:2002vw} to $(x(\bar{u}(x)+\bar{d}(x)$ and $x(s(x)+\bar{s}(x))$ respectively. The light blue dotted line is an CTEQ6.5s~\cite{Lai:2007vv} fit with a less constrained shape for the strange distributions relative to the light sea quarks. The blue band is the $\pm$ 1 $\sigma$ band of the strange quark distributions of the NNPDF2.3 fit~\cite{Ball:2012gg} which does not ab initio impose a shape on the parton distributions. Note the ATLAS result~\cite{Aaboud:2016btc} of eq. (27) shown as the green band suggests that by x of 0.023 $s+\bar{s} \sim \bar{u}+\bar{d}$ at a scale of 1.9 GeV$^2$.    
}
\label{fig:9}
\end{figure}

Another technique that is sensitive to the flavor of the quark is semi-inclusive deep inelastic scattering. Experimentally, hadron production in deep inelastic scattering is observed to factorize into scattering from an initial parton and fragmentation as the struck quark jet forms color neutral hadrons. For a hadron, $h$, that carries a fraction, $z$ (equals the energy of the hadron divided by the energy of the virtual photon, both in the lab frame), of the momentum of the struck quark
\begin{align}
\frac{d\s^h}{dx\,dQ^2\,dz} = \sum_{i=f,\bar{f}}\, K_i \, q_i(x,Q^2)\,D_i^h(z),
\end{align}
where $K$ is a kinematic factor containing the hard scattering cross section. The  expectation~\cite{Andersson:1998tv} is that if $z$ is sufficiently, but not too large, the most energetic (leading) hadron has a significant probability of containing a quark with the same flavor as the struck quark. This quark flavor retention was experimentally tested by the EMC collaboration in deep inelastic muon scattering~\cite{Albanese:1984nv} and is supported by models of fragmentation such as the Lund model~\cite{Andersson:1998tv}. Several criteria for the regions of kinematics where this assumption is valid have been proposed~\cite{Berger:1987zu,Boglione:2016bph,Mulders:2000jt}. Semi-inclusive DIS results from HERMES~\cite{Ackerstaff:1998} for the difference $\bar{d}-\bar{u}$ were shown in Fig. 3 and were in agreement with the Drell-Yan results.   This technique has been used by the HERMES~\cite{Airapetian:2004zf} and COMPASS~\cite{COMPASS:2010} collaborations to study the flavor dependence of the spin structure functions. Their results for polarized quark distributions are in reasonable agreement. Of course the fragmentation functions introduce new sources of uncertainty.  There have been several global analyses including de Florian {\it et al.}~\cite{deFlorian:2009vb}, Leader, Sidorov and Stamenov~\cite{Leader:2011tm}, and Ethier, Sato and Melnitchouk~\cite{Ethier:2017zbq}. Fig.~\ref{fig:8} illustrates the recent results from Ethier, Sato and Melnichouk. While the total spins carried by u and d quarks are well constrained, even the signs of the antiquark distributions are uncertain. Their result for the integral of the flavor difference in the sea quark spin distributions is $\int_0^1 dx \lf[\Delta \bar{u}(x) - \Delta \bar{d}(x)\rg] = 0.05 \pm 0.08$. This quantity will be discussed in the context of various models and calculations below.

HERMES~\cite{Airapetian:2013zaw} has also used semi-inclusive kaon production  to study the unpolarized strange quark distribution. Consider $Q(x) = u(x) + \bar{u}(x) + d(x) + \bar{d}(x)$ and $S(x) = s(x) + \bar{s}(x)$. For scattering from an isoscalar target like deuterium, the number of inclusive DIS events $N^{DIS}$, can be written in leading order as
\begin{align}
\frac{d^2\,N^{DIS}(x)}{dx\,dQ^2} = K_U(x,Q^2) \lf[5\,Q(x,Q^2) + 2\,S(x,Q^2)\rg],
\end{align}
where $K_U$ is a kinematic factor depending on the cross section and one has assumed charge symmetry, 
$u_p = d_n$, $d_p = u_n$, $\bar{u}_p = \bar{d}_n$ and $\bar{d}_p = \bar{u}_n$.

The number of charged kaons produced, $(N^K = N^{K^+} + N^{K^-})$,  is given by
\begin{align}
\frac{d^2N^K(x)}{dx\,dQ^2} &= K_U(x,Q^2) \no \\
&\hs*{0mm}
\times\lf[Q(x)\int dz\,D_Q^K + S(x)\int dz\, D_S^K(z)\rg].
\end{align}
If charge-conservation invariance is assumed in fragmentation, then there are only two functions describing the fragmentation involved.  $D_Q^K = 4\,D_u^K(z) + D_d^K(z)$ and $D_s^K \equiv 2\,D_s^K(z)$. If the fragmentation functions are well enough known, these two equations can be solved for $S(x)$.

Figure~\ref{fig:9} shows the HERMES results for $xS(x)$ compared to a CTEQ6 leading order fit for the strange distributions and the sum of the light antiquark distributions, $x\lf[\bar{u}(x) + \bar{d}(x)\rg]$~\cite{Pumplin:2002vw}. Also shown are a neural network fit~\cite{Ball:2012gg} and a CTEQ6.5S-0 fit where the shape of the S(x) is not constrained to be the same as that of the light quarks. The HERMES result has quite a different shape than the usual global fits, suggesting that there is little strange quark content for $x > 0.1$ and is more similar to the neural network result in this $x$ range. While this neural network fit was not consistent with the $W$ production results of Eq.(27), more recent NNPDF results~\cite{Ball:2016b} including the ATLAS W and Z production data do approach 1 at $x<0.01$ for the ratio of strange to light sea.  None of the semi-inclusive data were included in these unpolarized global fits. COMPASS also has measured the charged kaon multiplicities from $160\,$GeV muon scattering on deuterium~\cite{Adolph:2016bwc}. Both the summed charged kaon multiplicities and the ratio of $K^+$ to $K^-$ multiplicities are distinctly different from the HERMES results, while they agree for the ratio of charged pions in the region of overlap~\cite{Adolph:2016bga} but not the sum. Guerrero and Accardi~\cite{Guerrero:2018} suggest that hadron mass corrections may account for much of the discrepancy. If so, it is not yet known what effect this would have on the comparison of the polarized antiquark distribution results from the semi-inclusive analyses. In an independent analysis, Borsa, Sassot and Stratmann~\cite{Borsa:2017a} analyze the kaon multiplicities with simultaneous variation of the parton distributions and fragmentation functions and obtain strange quark densities close to the NNPDF 3.0 set. Still, at the present time, the differences in the kaon multiplicities from these two semi-inclusive DIS experiments and also the differences in inferred strange quark densities with the multi-muon neutrino data are not clearly understood. 

\begin{figure}[tbp]
\centering\includegraphics[width=\columnwidth]{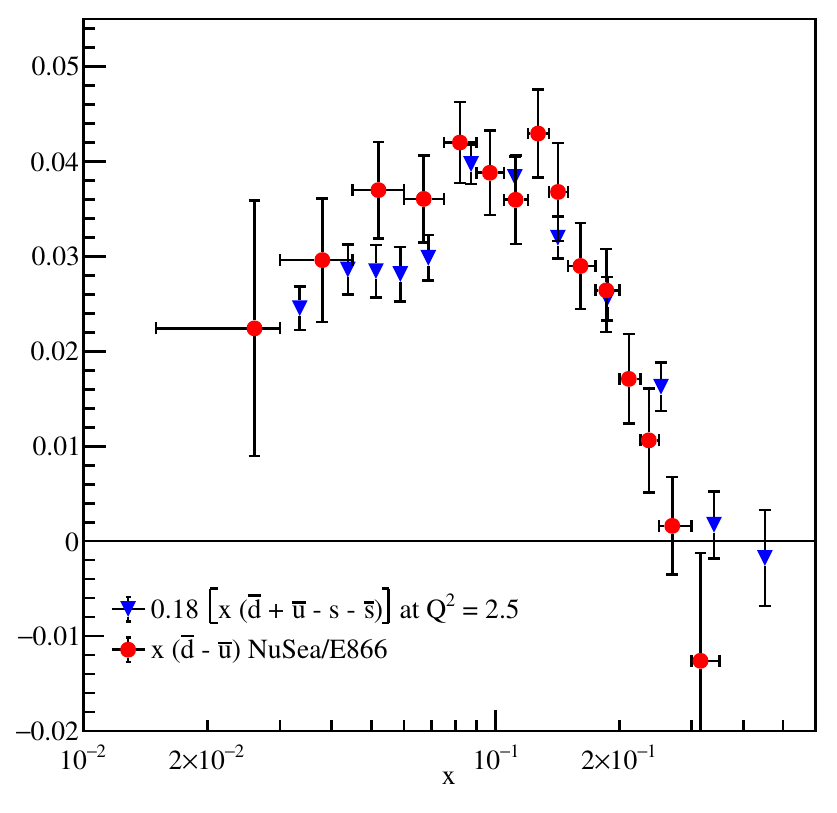}
\caption{$x(\bar{d}(x)-\bar{u}(x))$ vs $0.18 * x(\bar{d}+\bar{u}(x)-s(x)-\bar{s}(x))$ using NUSEA~\cite{Towell:2001nh} $\bar{d}-\bar{u}$ evaluated at 54 GeV$^2$, and HERMES~\cite{Airapetian:2013zaw} $s+\bar{s}$, and CTEQ6l1~\cite{Pumplin:2002vw} $\bar{u}+\bar{d}$, evaluated at 2.5 GeV$^2$.
}
\label{fig:10}
\end{figure}

 The combination of the HERMES and ATLAS data inspires a speculation that the strange quark distributions may be dominated by gluon splitting while the light antiquark distribution may have a substantial non-perturbative piece. Fig.~\ref{fig:10} shows the combination $0.18\,x\lf(\bar{u} + \bar{d} - s -\bar{s}\rg)$ using CTEQ6l1 leading order light quark distributions and the HERMES $S(x)$. Also shown is $x(\bar{d}-\bar{u})$ as determined from the Drell-Yan data (Fig.~\ref{fig:2} in a next-to-leading order analysis). The comparison is only qualitative since it involves results from different scales and from leading order and next-to-leading order analyses. However, the shapes of the two distributions are remarkably similar suggesting whatever the non-perturbative origin of the flavor asymmetry is also having a big effect on the total light quark sea at $x \gtrsim 0.07$. This similarity may provide a clue that pion degrees of freedom play a central feature in the explanation.

\subsection{Intrinsic Charm}
Just as production of strange mesons can be used to measure the intrinsic strangeness in the nucleon, production of charm is used to look for intrinsic $c\bar{c}$ components of the nucleon, that is charm at non-perturbative scales that is not produced by QCD evolution. Indeed, higher than expected production cross sections for charmed mesons in pp collisions led to the suggestion by Brodsky {\it et al.}~\cite{Brodsky:1980pb} of finite intrinsic charm. In neutral current deep inelastic scattering, the issue is to separate intrinsic charm contributions from the QCD process of photon-gluon fusion. The expectation is that due to the heavier charm quark mass, intrinsic charm would show up at high $x$. Experimentally, this has been studied by the EMC collaboration at CERN~\cite{Aubert:1981ix} and the ZEUS and H1 experiments at HERA~\cite{Abramowicz:1900rp} by detecting multi-muon events. There is some tension between these data sets. New data at high $x$ would be very welcome. One recent analysis~\cite{Jimenez:2015} that also included SLAC J$/\Psi$ data places upper limits on the average $x$ of intrinsic charm of $0.5\%$ and on the magnitude of the $c\bar{c}$ component of less that $1\%$. A next-to-leading order analysis by the NNPDF collaboration~\cite{Ball:2016} finds that for x<0.1 the data are consistent with only perturbative charm which vanishes at $Q \approx 1.6$ GeV, but  an “intrinsic” large x component is required, peaking at $x \sim 0.5$, carrying $0.7 \pm 0.3 \%$ of the nucleon's momentum. In an updated next-to-next-to leading order analysis~\cite{Ball:2016b} this is reduced to $0.26 \pm 0.42 \% $ at a scale of the charm quark mass. The sensitivity of this result to inclusion of the EMC data or an assumption of perturbatively generated charm is discussed. They note that at high x even a small non-pertubative component can lead to a significant impact for specific LHC cross sections.

The model of Brodsky {\it et al.} ~\cite{Brodsky:1980pb} will be applied to the lighter sea quarks below.

\section{Calculations and Models of the Sea Distributions}

There have been a number of excellent reviews of the sea quark distributions over the years. These include Kumano~\cite{Kumano:1997cy}, Garvey and Peng~\cite{Garvey:2001yq} and Chang and Peng\cite{Chang:2014jba}. A recent (October 2017) Institute for Nuclear Theory Workshop, ``The Flavor Structure of the Nucleon Sea'', provides an excellent overview of the current theoretical situation. The presentations at this workshop can be found at~\cite{INT:17}. The literature is sufficiently large that only a sampling of the theoretical papers on each approach are discussed here. Generally, the references of the cited papers or the earlier reviews provide more comprehensive lists of similar approaches. 

\begin{figure}[tbp]
\centering\includegraphics[width=\columnwidth]{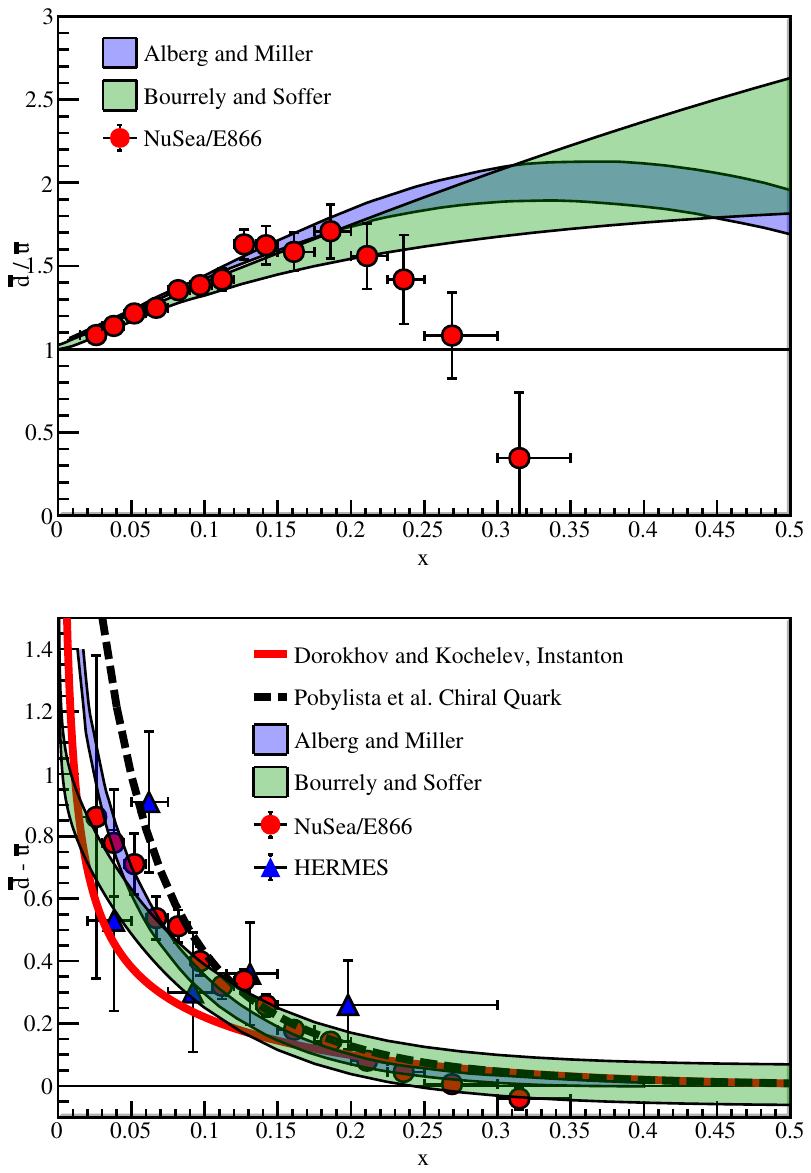}
\caption{The upper graph shows the NUSEA~\cite{Towell:2001nh}  $\bar{d}/ \bar{u}$ data compared to the hadron model calculation of Alberg and Miller~\cite{Alberg:2017ijg}  and the statistical parton distribution fit of Bourrely and Soffer\cite{Bourrely:2015kla}.  The lower plot shows $\bar{d} -\bar{u}$ from NUSEA~\cite{Towell:2001nh}  and HERMES~\cite{Ackerstaff:1998} along with an instanton model~\cite{Dorokhov:1993}, chiral quark solition model~\cite{Pobylitsa:1999}, the meson cloud model of Alberg and Miller~\cite{Alberg:2017ijg}  and the statistical parton distribution fit\cite{Bourrely:2015kla}.
}
\label{fig:11}
\end{figure}
\subsection{Impact of the Pauli Principle}

While the antisymmetry of fermions in a quantum mechanical wave function is a fundamental feature of quantum mechanics, it barely receives attention in relativistic bound states with multiple Fock-state components. Naively, there are 6 $s$-wave states available for each flavor quark, with 2 spin states and 3 color states. In a proton two of these are filled by the $u$ valence quarks and one for the $d$ valence quarks, so one might expect $5/4$ as many $d\bar{d}$ excitations as $u\bar{u}$ excitations at some low scale. This value is much smaller than the observed peak $\bar{d}/\bar{u}$ ratio in Fig.~\ref{fig:2}. In the large $N_c$ approximation of a large number of colors, there would be no preference.

Steffens and Thomas~\cite{Steffens:1996bc} find that the interference between the sea quarks generated by gluon emission and the remaining quarks in the nucleon hide the effects of the Pauli principle and, indeed, can lead to an excess of $\bar{u}$ compared to $\bar{d}$ though the effect is numerically very small. They point out that similar issues occur in lattice QCD, and that due to the Pauli principle, neither the connected insertions or the disconnected insertions are physically meaningful alone.

As one moves to higher scales, there is no experimental evidence for corrections to the DGLAP evolution from Pauli blocking.

Bourrely, Soffer and Buccela~\cite{Bourrely:2015kla,Bourrely:2005kw} treated the individual helicity parton distributions for each flavor as finite temperature Fermi-Dirac distributions at a scale, $Q_0^2$ of $\sim 1$ GeV$^2$ ($3\,$GeV$^2$ in the 2nd reference). For the valence quarks, the ansatz is a Fermi-Dirac function depending on helicity, $h$, and a helicity-independent diffractive contribution
\begin{align}
x\,q^h(x,Q_0^2) = \frac{A_q\,X_{0q}^h\,x^b}{{\rm \exp}\lf[\frac{x - X_{0q}^h}{\bar{x}}\rg] + 1}
+ \frac{\tilde{A}_q\,x^{\bar{b}}}{{\rm \exp}\lf[\frac{x}{\bar{x}}\rg] + 1},
\end{align}
where $X_{0q}^h$ is a constant, which plays the role of the thermodynamic potential and $\bar{x}$ is the universal temperature. The chiral structure of QCD gives the constraints that
\begin{align}
X_{0q}^h &= -X_{0\bar{q}}^{-h}, & X_{0g} = 0.
\end{align}
For the light and strange antiquarks
\begin{align}
x\,\bar{q}^h(x,Q_0^2) = \frac{\bar{A}_q\,(X_{0q}^h)^{-1}\,x^{b_{\bar{q}}}} {{\rm \exp}\lf[\frac{x + X_{0q}^{-h}}{\bar{x}}\rg] + 1}
+ \frac{\tilde{A}_q\,x^{\bar{b}}}{{\rm \exp}\lf[\frac{x}{\bar{x}}\rg] + 1}.
\end{align}
%
%
%
%
The $c$, $b$, and $t$ distributions are set to 0 at the initial scale.

The results of their fits to DIS data are that the six potentials satisfy
\begin{align}
X_{0u}^+ > X_{0u}^- \approx X_{0d}^- > X_{0d}^+ \gg X_{0s}^- > X_{0s}^+ ,
\end{align}
leading to the predictions that $\bar{d}(x) > \bar{u}(x)$ and $\bar{d}(x) - \bar{u}(x) \approx \Delta \bar{u}(x) - \Delta \bar{d}(x)$. At high $x$, the ratio of $\bar{d}/\bar{u}$ flattens out at about $2.5$ in their model. Their results are shown as the green bands in the upper panel of Fig.~\ref{fig:11} and as the dashed-dot curve in the lower panel of Fig.~\ref{fig:11}. 

Zheng, Zhang and Ma~\cite{Zhang:2001plb} argue that the principle of detailed balance in gluon splitting and recombination naturally leads to a sea quark asymmetry. For example, there are three ways a  $|uudu\bar{u} \rangle $ can transition to a $|uudg \rangle$ but only two ways a $|uudd\bar{d} \rangle $ can transition to a  a $|uudg \rangle$. Assuming a statistical ensemble of Fock states and a normalization condition, they predict
\begin{align}
  \int_0^1 dx (\bar{d}(x) - \bar{u}(x)) = 0.13 ,
\end{align}
with no free parameters, in remarkable agreement with the experimental result. In a slightly more general model~\cite{Zhang:2010prd}, they predict flavor asymmetries for other octet baryon states and kaons.

\begin{figure}[tbp]
\centering\includegraphics[width=\columnwidth]{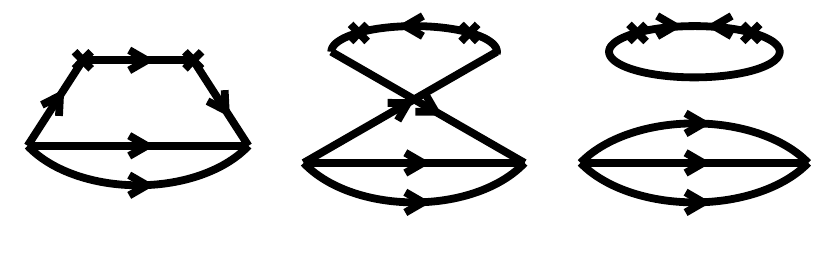}
\caption{Lattice QCD topologically distinct connected and disconnected diagrams. In between the currents at the X's, a) shows connected valence and sea quarks, b) connected sea antiquarks, and c) disconnected sea quarks and antiquarks. 
}
\label{fig:12}
\end{figure}

\subsection{Lattice QCD}
The only ab initio technique for solving the structure of the nucleon is lattice QCD, where space time is discretized for numerical solutions to the QCD Lagrangian. This subject has recently been reviewed in Lin {\it et al.}~\cite{Lin:2017snn}. A primary difficulty is that lattice calculations are done in Euclidian space while parton distributions require light-cone dynamics. For many years lattice calculations of parton distributions were based on the operator product expansion as forward nucleon matrix elements of local twist-2 operators which are directly related to moments of the parton distribution functions~\cite{Detmold:2001jb}.
\begin{align}
\lf<x^n\rg> = \int_0^1 dx\, x^n\lf[q(x,Q^2) + (-1)^{n+1}\,\bar{q}(x,Q^2)\rg].
\end{align}
These calculations are limited to the first few moments as noise increases for the higher moments. In practice it was recognized that one could not reliably obtain the x dependence by this technique. The matrix elements involve both connected and disconnected diagrams (Fig~\ref{fig:12}). The evaluation of disconnected diagrams is considerably more difficult, and only recently have they been included. Liu~\cite{Liu:2017lpe} argues that the disconnected diagrams are flavor independent and derives separate evolution equations for the connected and disconnected diagrams. However, Steffens and Thomas~\cite{Steffens:1996bc} suggest the effects of the Pauli principle could still lead to a flavor asymmetry here also. Lattice calculations of the disconnected diagrams with realistic pion masses are only now reaching the point where this can be tested.

A promising new technique to directly calculate parton physics on the lattice is large-momentum effective theory [LMET]~\cite{Ji:2013dva,Ji:2017oey}. First results on the flavor structure are beginning to become available, and the systematic errors of this new technique are being evaluated. Lin {\it et al.}~\cite{Lin:2015} report an asymmetry in the integral of $(\bar{d}-\bar{u}) $ of $0.14 \pm 0.05$ and an asymmetry in the integral of the polarized sea quark asymmetry, $\Delta \bar{u} - \Delta \bar{d}$ of $0.24 \pm 0.06$. In a later conference proceedings, Lin~\cite{Lin:2016} reports numbers for the integral of $\bar{d}-\bar{u} $ of 0.13 $\pm$ 0.07 and $\int_{0.08}^1 ( \Delta \bar{u} - \Delta \bar{d}) dx$ = 0.14 $\pm$ 0.09.  The error bars are still sizable, but the first results might suggest $\Delta \bar{u} - \Delta \bar{d} > \bar{d}-\bar{u} $ as suggested by models where the number of colors is large but they are 2 $\sigma$ away from the global fit result for the polarized difference of Ref ~\cite{Ethier:2017zbq}. The second result of near equality of the flavor and spin integrals is near the statistical model expectations. Another example of extremely promising work comes from the publications of Alexandrou {\it et al.} also with physical values of the pion masses (~\cite{Alexandrou:2018a}~\cite{Alexandrou:2017a}~\cite{Alexandrou:2015a} and references therein). In their most recent publication ~\cite{Alexandrou:2018b}, they validate the methodology of the LMET approach and identify the remaining systematic effects. Examples of the current state of the lattice work adapted from ~\cite{Lin:2017snn} are shown in Figure 13. There remain large uncertainties and some disagreement with the global fit results, for example in $x(\Delta u- \Delta d)$ (not shown) where the experimental errors are considerably smaller. In another significant step, transversity parton distributions can now be calculated~\cite{Alexandrou:2018c}. Based on these recent works, it is anticipated that there should be rapid progress in reducing the lattice uncertainties in the next few years, and the lattice soon may be providing extremely important insights.

\begin{figure}[tbp]
\centering\includegraphics[width=\columnwidth]{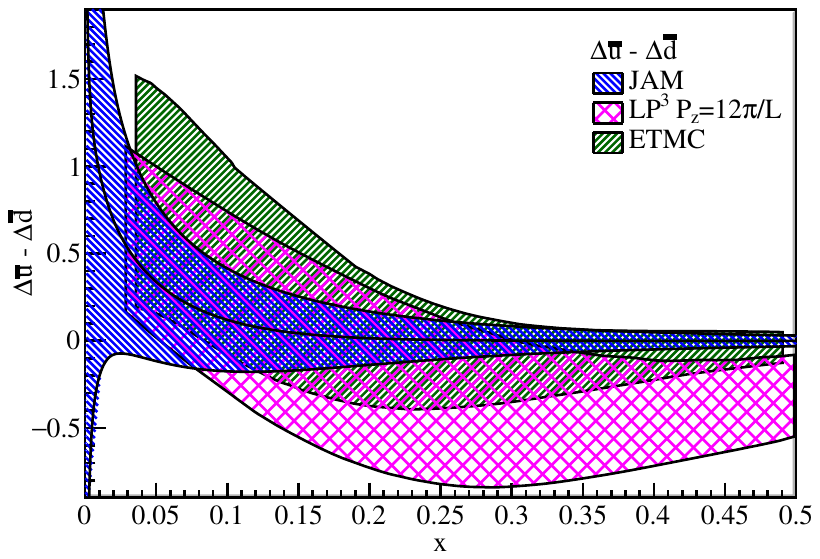}
\caption{Examples of recent lattice QCD results for isovector antiquark parton spin distributions, $(\Delta \bar{u}(x) - \Delta \bar{d}(x))$, compared to the JAM15 global fit~\cite{Ethier:2017zbq} (blue).  The magenta band is the LP3 lattice calculations of Lin {\it et al}~\cite{ Lin:2017ipdf} and the green band is the ETMC lattice calculations of Alexandrou {\it et al.}~\cite{ETMC:2017}. Only statistical errors are shown. Lin {\it et al.}~\cite{Lin:2017snn} note that the small x region, of primary interest for the sea quarks, can suffer additional systematics due to the limited nucleon boost momentum.  
}
\label{fig:13}
\end{figure}

\subsection{Instantons}
Instantons are topological, non-trivial, four-dimensional gluon field configurations that solve the U(1) problem and contribute to path integrals in QCD. In the 1980's a picture of the QCD vacuum emerged as an dilute liquid of well localized topological fluctuations. The small size of the instantons was given as the reason the chiral symmetry breaking scale was so large, $\sim$ 1 GeV, why pions are so light, and why glueballs are heavy~\cite{Schaffer:1997}. Early quenched lattice calculations gave some support to this picture.  More recent lattice calculations suggest a much more complicated gluon structure, and in the end, lattice simulations should determine this. Forte and Shuryak~\cite{Forte:1991} showed that instanton-anti-instanton pairs contribute to the isosinglet axial current in a polarized proton. The 't Hooft effective Lagrangian~\cite{Hooft:1976} couples  $\bar{u}_R u_L \bar{d}_R d_L$ and also $\bar{u}_L u_R \bar{d}_L d_R$, so the interaction with an instanton can change, for example, a $u_L$ into a $u_R \bar{d}_R d_R$, creating an excess of $\bar{d}$. Dorokhov and Kochelev~\cite{Dorokhov:1991}~\cite{Dorokhov:1993} demonstrated that at large x the ratio of $\frac{\bar{d}}{\bar{u}}$ goes to 4 (much larger than what is seen so far in Fig. 2) and, indeed, $\frac{\bar{u} + \bar{d}}{2 \bar{s}}$ goes to 1. They also predict that $\int (\Delta \bar{u}-\Delta \bar{d}) dx = \frac{5}{3} \int (\bar{d}-\bar{u})dx$.  In the first publication they estimated the integrated antiquark difference to be $0.24 \pm 0.1$. In the second publication, they make assumptions about
the x dependence of the parton distribution functions, fit the integrated antiquark difference, and make predictions for the instanton contribution to the polarized distributions and the Drell-Yan asymmetry. The latter does not match the existing data very well.

\subsection{Hadron Models}
There is a long history of considering the impact of meson-baryon fluctuations on physical baryon properties. That a neutron could fluctuate into a proton and a $\pi^-$ is a simple explanation of why the neutron charge density appears to be positive in the center and negative at longer distances. Sullivan~\cite{Sullivan:1971kd} first considered the deep inelastic scattering from the pion cloud of the proton. In the late 1970's and early 1980's chiral bag models showed that pion fields were required to preserve chiral symmetry at the bag boundaries. Thomas~\cite{Thomas:1983fh} investigated the impact of the pion cloud on SU(3) breaking, essentially predicting $\bar{d} - \bar{u} > 0$ with about the right magnitude. Signal and Thomas~\cite{Signal:1987gz} extended the calculations to the strange sea and, for example, Cao and Signal~\cite{Cao:2003} addressed the non-perturbative structure of the polarized sea .

It is easy to see that if the physical proton is made up of a bare proton and a bare nucleon plus pion cloud
\begin{align}
|\vect{p}\big> &= \a|p\big> + \b\lf[c_{1/2,0}|p\,\pi^0\big> + c_{-1/2,1}|n\,\pi^+\big>\rg], \no \\
&= \a|uud\big> + \b\lf[c_{1/2,0} |uud + \frac{u\bar{u}+d\bar{d}}{\sqrt{2}}\big> + c_{-1/2,1}|udd+u\bar{d}\big>\rg],
\end{align}
where $c_{1/2,0}$ and $c_{-1/2,1}$ are the isospin Clebsch-Gordan coefficients 
$\lf<1/2,1/2,1,0|1/2,1/2\rg> = -\sqrt{\frac{1}{3}}$ and $\lf< 1/2,-1/2, 1, 1|1/2,1/2\rg> = \sqrt{\frac{2}{3}}$ 
respectively. The high energy convention for the sign of $T_z = \frac{1}{2}$ for the proton is used. In higher $x$ regions where the gluon splitting generated sea might be negligible, this predicts $\bar{d}/\bar{u}=5$, much higher than the experimental ratio. If one adds a Delta resonance component
\begin{align}
&\g\Bigg[
  c_{3/2,-1}|uuu + \bar{u}d\big> \no \\
&\hs{10mm}
+ c_{1/2,0} |uud + \frac{u\bar{u}+d\bar{d}}{\sqrt{2}}\big > 
+ c_{-1/2,1}|udd + u\bar{d}\big> 
\Bigg]
\end{align}
then the appropriate Clebsch-Gordan coefficients squared $\lf<3/2,m_Z^\Delta,1,m_Z^\pi|1/2,1/2\rg>^2$ are $1/2$, $1/3$, and $1/6$. If $\g^2$ were much larger than $\b^2$ and the gluon splitting generated sea were negligible then $\frac{\bar{d}}{\bar{u}} = 1/2$. However, this does not seem to be reasonable physically. Peng {\it et al.}~\cite{Peng:1998pa} used the E886 results and estimates that $\beta^2 \approx 2\,\g^2$ to infer that $\beta^2 = 0.20 \pm 0.04$.

A distinctive feature of these pion models is that all the antiquarks are contained in spin-zero pions and so cannot have a preferred orientation. Therefore, $\Delta \bar{u} = \Delta \bar{d} = \Delta \bar{u} - \Delta \bar{d} = 0$. On the other hand, since the nucleon and pion must be in a relative $P$ wave to conserve parity and angular momentum, one might expect the antiquarks to reveal the presence of orbital angular momentum, for example, through a non-zero Sivers function (for example, \cite{Brodsky:2002} \cite{Lu:2007}).

From here, one could add more baryon and meson states. Once vector mesons are included, the net spin of the antiquarks can be non-zero. Similarly, with the inclusion of strange baryons and mesons, one can try to calculate properties of the strange sea. The primary issues are where to truncate the hadronic expansion and how to properly include meson-nucleon form factors in frame independent manner that incorporates experimental input. Many of these issues have been handled by Alberg and Miller~\cite{Alberg:2012wr}. Recent predictions of the Alberg and Miller calculations in the context of chiral light front perturbation theory~\cite{Alberg:2017ijg} are shown in Fig.~\ref{fig:11}. New data expected soon at higher x should be decisive for this approach. Clearly, if higher precision data confirm the rapid drop at x$\sim$ 0.3, it will be inconsistent with this picture.

A further puzzle for a hadronic fluctuation description comes from leading proton and neutron production in deep inelastic scattering at HERA where a nucleon with near beam velocity is detected at zero degrees~\cite{Chekanov:2007}~\cite{Chekanov:2009}. These data are often interpreted in terms of a pion-nucleon fluctuation, and indeed are used to extract the structure function of the pion (See, for example, Levman~\cite{Levman:2002}). However the yield of leading protons is twice that of leading neutrons, in contrast to the expectation from the isospin Clebsch-Gordan coefficients above. In some models, Pomeron and isoscalar Reggeon exchange dominate over much of the measured region~\cite{Szczurek:1998}. 

\subsection{Chiral Effective Theory}
Another approach uses chiral effective field theory. (See, for example, ~\cite{Detmold:2001jb} which is extended to the strange quark sea in, for example, Wang {\it et al.}~\cite{Wang:2016ndh}~\cite{Wang:2016prd} and references therein.) Thomas, Melnitchouk and Steffans~\cite{Thomas:2000ms} showed that in chiral expansions of the moments of strange-quark distributions, the coefficients of leading non-analytic terms in the kaon mass are model independent and can only arise from pseudoscalar loops. Chiral effective theory starts with the most general effective Lagrangian for the interaction of an octet of baryons through pseudoscalar fields. These include the so-called ``kaon rainbow,''  ``kaon bubble,'' ``hyperon rainbow,''  ``kaon tadpole,'' and Kroll-Ruderman diagrams. In reference ~\cite{Thomas:2000ms} the integral of the leading non-analytic contribution to the integral of $(\bar{d}(x)-\bar{u}(x))$ was estimated be about 0.2, most of which comes from the pion-nucleon terms. For the strange sea, the integral of $x( s(x) - \bar{s}(x))$ ranges from 0.4-1.1 $ \times 10^{-3}$ to be compared with the experimental number given in Eq. 32.     

One novel consequence of this approach is that the parton distributions contain a delta function contribution at $x=0$, implying that the total integral of $s(x)$ or $\bar{s}(x)$  is experimentally inaccessible. On the other hand the x weighted integral is well defined since the delta function contributions at x=0 vanishes.

\subsection{Quarks and Mesons}
An intermediate picture is to envision that the mesons couple directly to the valence quarks. The coupling is governed by similar isospin Clebsch-Gordan coefficients
\begin{align}
|u\big> &\to \b\lf[c_{1/2,0}|u\,\pi^0\big> + c_{-1/2,1}|d\,\pi^+\big>\rg], \\
|d\big> &\to \b\lf[c_{-1/2,0}|d\,\pi^0\big> + c_{1/2,-1}|u\,\pi^-\big>\rg].
\end{align}
Here the isospin coupling would give a ratio of $\bar{d}/\bar{u} = 11/7$, close to peak of the experimental results.

\subsection{Chiral Soliton Models}
In the limit of a large number of colors, large $N_c$, QCD becomes equivalent to an effective theory of mesons, and baryons appear as solitons. The calculations are typically based on an effective action derived from the instanton vacuum of QCD~\cite{Diakonov:1987ty,Diakonov:1997vc,Diakonov:1997sj,Pobylitsa:1999}. Numerical calculations of the $x$ dependence of $\bar{d} - \bar{u}$ are shown in Fig.~\ref{fig:11}~\cite{Pobylitsa:1999}. At high $x$ the ratio of $\bar{d}/\bar{u}$ is predicted to be $11/7$ as in the pion+valence quark picture discussed above. As in the instanton picture, there is a close connection between the unpolarized and polarized isovector sea contributions with
\begin{align}
\bar{d}(x) - \bar{u}(x) \approx \frac{3}{5}\lf[\Delta \bar{u}(x) - \Delta \bar{d}(x)\rg].
\end{align}

In a similar approach Wakamatsu and Watabe~\cite{Wakamatsu:2010prd} obtain a slightly more complicated relation between the antiquark flavor and spin differences which they parameterize as
\begin{align}
  [\Delta \bar{u}(x) - \Delta \bar{d}(x)] = 2.0 x^{0.12} [\bar{d}(x) - \bar{u}(x)].
\end{align}


\subsection{Five-Quark Fock States}
Brodsky {\it et al.} proposed a phase-space-inspired distribution for a five-quark Fock state in a proton at some low scale as
\begin{align}
P(x_1,\ldots,x_5) = N_5\,\delta\!\lf(1 - \sum_{i=1}^5x_i\rg)
\lf[m_p^2 - \sum_{i=1}^5\,\frac{m_i^2}{x_i}\rg]^{-2},
\end{align}
where $m_p$ is the proton mass and $m_i$ is the mass of quark $i$. The delta function ensures momentum conservation. They were focused on $c\bar{c}$ states. Chang and Peng~\cite{Chang:2014lea} extended this analysis to the lighter quarks, first by fitting the strange sea by evolving the distribution from an initial scale to the $Q^2$ of the HERMES results discussed above, and then determining the u and d quark sea using the combination $\bar{u} + \bar{d} - s - \bar{s}$ based on the HERMES and NUSEA results. The results are sensitive at the 20-30\% level to the choice of initial scale of $0.5$ or $0.3\,$GeV. Typical results fitting the 2014 HERMES analysis and the NUSEA results gives probabilities of $u\bar{u}$, $d\bar{d}$, and $s\bar{s}$ of $0.19$, $0.31$ and $0.11$ respectively. The results are also quite sensitive to the choice of kaon fragmentation functions. They conclude the HERMES results do not exclude the existence of an intrinsic strange-quark component of the nucleon sea.

\section{Prospects for Future Work}

It is anticipated that the first new results, expected very soon, will be SeaQuest data of 120 GeV proton induced Drell-Yan measurements on hydrogen, deuterium and several nuclear targets. These data will extend the x range of $\frac{\bar{d}}{\bar{u}}$ to x of 0.4-0.5 and will decisively confirm or refute the suggestion of a decrease in  $\frac{\bar{d}}{\bar{u}}$ above x of 0.2,

The Jefferson Lab 12 GeV upgrade will provide extremely high luminosity polarized deep inelastic and semi-inclusive deep inelastic data for x> 0.1, though at modest Q$^2$.  Figure 14 illustrates the projected sensitivity of the measurement of $\Delta \bar{u} - \Delta \bar{d}$ from an approved CLAS-12 measurement(E12-09-007)~\cite{Hafidi:2012}. High statistics data will also be obtained in semi-inclusive kaon production. If the issues with hadron mass corrections can be satisfactorily understood, such data should lead to better strange quark distributions.

RHIC experiments~\cite{Aschenauer:2016rcq} have accumulated data with a total luminosity of about 400 pb$^{-1}$  with longitudinally polarized protons at $\sim$ 500 GeV center of mass energy and $\sim$ 85 pb$^{-1}$ at 200 GeV center of mass energy. Final results for W production and the spin carried by the sea quarks should be available soon and are expected to place better constraints on $\Delta \bar{u}$ and $\Delta \bar{d}$. Runs in 2015 and 2017 focused on transversely polarized protons and orbital angular momentum related to the Sivers function. First results with 25 pb$^{-1}$, albeit with sizable statistical errors, are consistent with the expected sign flip of the Sivers function between deep inelastic scattering and vector boson production~\cite{Adamczyk:2016}. These will be considerably improved with the 2017 data. For the next several years, RHIC's focus will be on the beam energy scan. Further polarized proton running is not currently anticipated before 2021~\cite{Aschenauer:2016rcq}. 

The COMPASS experiment in 2017 acquired more semi-inclusive DIS data. In 2018 they plan to continue pion-induced Drell-Yan to study the Sivers function of the valence quarks before the SPS shuts down in 2019-2020. They are proposing~\cite{Badelek:2018} a 2021 run with muons on a transversely polarized deuterium target to improve the measurements of the transversity distribution h$_1^d$ and the nucleon tensor charge and transverse momentum distributions. In the longer term, plans are in progress for a proton radius measurement and radiofrequency-separated hadron beams, for, among other physics, Drell-Yan measurements with beams of kaons and anti-protons.    

The impact of LHC data is already clearly seen in the discussion of the strange quark sea. A recent study by Khalek et al.~\cite{Khalek:2018a} concludes that High-Luminosity LHC measurements, planned for the middle of the 2020's, can reduce the parton distribution functions uncertainties by factors of 2-5 depending on the $x$ range and specific channels. 

In the longer term, a high-luminosity polarized electron-ion collider would provide definitive information about the flavor dependence of the spin of the quarks and gluons in the proton and their orbital angular momentum.  The need for such a facility was a major recommendation of the 2015 Nuclear Science Advisory Committee Long Range Plan. 

\begin{figure}[tbp]
\centering\includegraphics[width=\columnwidth]{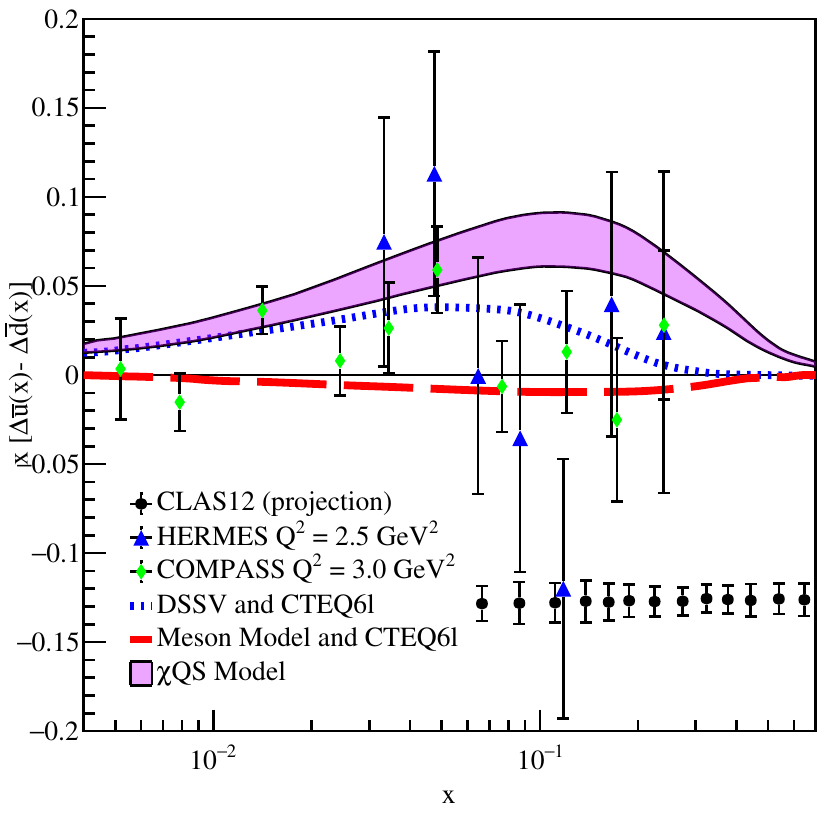}
\caption{Projected JLAB uncertainties for a semi-inclusive DIS measurement of $x(\Delta \bar{u} - \Delta \bar{d} )$ compared to HERMES~\cite{Airapetian:2004zf} and COMPASS~\cite{COMPASS:2010} data, an early global fit~\cite{deFlorian:2009vb}, another chiral quark soliton~\cite{Dressler:2000} model and another meson cloud model~\cite{Cao:2003}. 
}
\label{fig:15}
\end{figure}

\section{Summary\label{Conclusion}}

While the large contribution to the sea resulting from gluon splitting is well described at low x, the non-perturbative features of the sea are an essential aspect of proton structure that is still not understood. The trend of the NUSEA data to suggest that $\bar{d}/\bar{u}$ decreases rapidly above $x \sim 0.2$ and possibly becomes less than 1 at higher x does not seem to be consistent with any model. Admittedly, the error bars grow large at higher x. The SeaQuest experiment should provide higher statistics Drell-Yan measurements in this x range in the very near future. Precise semi-inclusive DIS measurements at Jefferson Lab and W$^{+/-}$ production at RHIC should sharpen the comparison of $\bar{d}-\bar{u}$ with the polarized sea $\Delta \bar{u}-\Delta \bar{d}$.  JLAB data should also shed light on the third puzzle of the x dependence of the strange sea distributions. High statistics LHC data will also contribute significantly to constraining the sea quark distributions. At the same time it appears that lattice results may become decisive in the near future. This combination of new experimental results and advances in theory give the authors confidence that the origin of a non-perturbative sea of the proton can become a solved problem in the next few years.

In addition to the fundamental insight into hadron structure provided by the proton sea, at the highest scales of discovery in proton-proton colliders like the Large Hadron Collider, the cross sections for  quark-antiquark coupling to new particles such as heavy Z's or W's depend directly on these non-perturbative features of the antiquark distributions at higher x. If $\bar{u}$ is indeed greater than $\bar{d}$ at high $x$ values, then  $Z^{\prime}$ production is favored over $W^{\prime}$ production in a pp collider, while if $\bar{d}$ is greater than $\bar{u}$, $W^{\prime}$ production is favored. Such considerations can affect the production yield on the same order as factors of 2-4 in luminosity of the LHC in the search for new physics. 

\begin{acknowledgments}
The authors would like to thank Ian Cloët for many helpful discussions. This work was supported by the U.S. Department of Energy, Office of Science, Office of Nuclear Physics, contract no. DE-AC02-06CH11357; and Laboratory Directed Research and Development (LDRD) funding from Argonne National Laboratory, project no. 2016-098-N0 and project no. 2017-058-N0. 
\end{acknowledgments}
\bibliographystyle{myapsrev4-1}
\bibliography{seaquarks}

\end{document}